%% file: sample-ccs2026-MTLS-minor-revision.tex
\documentclass[sigconf,nonacm]{acmart}
\AtBeginDocument{%
  }

\setcopyright{acmlicensed} 
\copyrightyear{2018} 
\acmYear{2018} 
\acmDOI{XXXXXXX.XXXXXXX} 
\acmConference[Conference acronym 'XX]{Make sure to enter the correct
  conference title from your rights confirmation email}{June 03--05,
  2018}{Woodstock, NY}  
\acmISBN{978-1-4503-XXXX-X/2018/06}  

\usepackage{csquotes}              
\usepackage{dirtytalk}             
\usepackage[all]{nowidow}          

\emergencystretch=\maxdimen
\let\savedsay\say
\renewcommand{\say}[1]{\savedsay{\textit{#1}}}

\usepackage[table]{xcolor}

\definecolor{citeorange}{HTML}{FF8200}
\definecolor{utkorange}{HTML}{FF8200}
\definecolor{quote}{rgb}{0.99, 0.99, 0.99}
\definecolor{bar}{rgb}{0.5, 0.5, 0.5}

\usepackage{booktabs}              
\usepackage{tabularx}              
\usepackage{multirow}              
\usepackage{array}                 
\usepackage{flafter}               
\usepackage{placeins}              

\newcolumntype{L}{>{\raggedright\arraybackslash}X}

\usepackage{listings}
\usepackage{xfp}                   

\newcommand{\totalN}{46}           
\newcommand{\pct}[1]{\fpeval{round(100*#1/\totalN, 0)}\%}

\usepackage{mdframed}
\usepackage{framed}
\usepackage[most]{tcolorbox}

\newcommand{\longsay}[2]{%
  \def\FrameCommand{%
    \hspace{2pt}%
    {\color{bar}\vrule width 2pt}%
    {\color{quote}\vrule width 4pt}%
    \colorbox{quote}%
  }%
  \MakeFramed{\advance\hsize-\width\FrameRestore}%
  \begin{list}{}{%
    \setlength{\topsep}{0pt}
    \setlength{\leftmargin}{0pt}
    \setlength{\rightmargin}{0pt}
  }
  \item[]
    \say{#1} (#2)%
  \end{list}%
  \endMakeFramed%
}

\newcounter{takeawaycnt} 
\newtcolorbox{takeawaybox}{
    enhanced,
    breakable,
    rounded corners,
    colback=gray!10,        
    colframe=gray!10,       
    borderline west={4pt}{0pt}{utkorange},
    boxrule=0pt,
    sharp corners=northeast,
    sharp corners=southeast,
    boxsep=3pt,
    left=5pt,
    right=2pt,              
    top=2pt,                
    bottom=2pt
}

\newenvironment{mytakeaway}
    {%
    \begin{takeawaybox}
    \textbf{Takeaway \thetakeawaycnt:} %
    \stepcounter{takeawaycnt}%
    }
    {%
    \end{takeawaybox}
    }

\usepackage{xurl}                  
\usepackage{hyperref}

\hypersetup{
  colorlinks=true,
  linkcolor={green!80!black},
  citecolor=citeorange,
  urlcolor={blue!70!black}
}

\begin{document}


\title[``Setting up TLS authentication was hell'']{``Setting up TLS authentication was hell'': A Usability Study of Client Certificate Authentication}

\thanks{\noindent\textcolor{red}{\textbf{A shortened version of this paper appears in the Proceedings of the 2026 ACM SIGSAC Conference on Computer and Communications Security (ACM CCS 2026). }}}



\author{Abubakar Sadiq Shittu}
\orcid{0000-0002-4699-3015}
\affiliation{%
  \institution{University of Tennessee}
  \city{Knoxville}
  \state{TN}
  \country{USA}%
}
\email{ashittu@vols.utk.edu}

\author{Clay Shubert}
\affiliation{%
  \institution{University of Tennessee}
  \city{Knoxville}
  \state{TN}
  \country{USA}%
}
\email{cshubert@alum.utk.edu}

\author{John Sadik}
\orcid{0009-0007-8988-5501}
\affiliation{%
  \institution{University of Tennessee}
  \city{Knoxville}
  \state{TN}
  \country{USA}%
}
\email{jsadik@vols.utk.edu}

\author{Scott Ruoti}
\orcid{0000-0002-6917-4186}
\affiliation{%
  \institution{University of Tennessee}
  \city{Knoxville}
  \state{TN}
  \country{USA}%
}
\email{ruoti@utk.edu}


\input{sections/0.abstract.tex}





\maketitle

\input{sections/1.intro.tex}

\input{sections/2._background_and_related.tex}

\input{sections/2._rw.tex}

\input{sections/3._participants.tex}

\input{sections/4._setup_m_r.tex}
\input{sections/5._usage_m_r.tex}
\input{sections/6._setup2_m_r.tex}

\input{sections/7._understanding.tex}
\input{sections/8._limitation.tex}
\input{sections/9._discussion.tex}

\begin{acks}

We appreciate the anonymous reviewers for their constructive suggestions and guidance.

\end{acks}

\bibliographystyle{ACM-Reference-Format}
\bibliography{reference}

\appendix 

\section{Open Science}
\label{sec:open_science}

To protect participant privacy and comply with IRB requirements, we do not release raw reflections, server logs, or signed certificates, as these could deanonymize participants. However, we release other study materials at \href{https://anonymous.4open.science/r/mtls_usability/README.md}{\textcolor{blue}{\nolinkurl{https://anonymous.4open.science/r/mtls_usability/README.md}}}. These materials include the recruitment email used to invite students to consent, the IRB-approved consent form, both assignments, and the qualitative codebook. We also provide artifacts to support mTLS deployment for both clients and servers, along with guidance for resolving common setup issues.

\input{sections/ethics.tex}


\input{sections/supp.tex}

\end{document}

%% file: sections/0.abstract.tex
\begin{abstract}

\emph{``Cryptography turns a security problem into a key management problem''}~\cite{aumasson_murphy}. Despite decades of research effort towards usable key management, it remains unclear whether key management issues are inherent to every cryptographic system or merely artifacts of specific designs. To investigate, this paper presents a user study of mutual TLS (mTLS) usability, tracking 46 senior and graduate computer science students—highly technical users who configured client certificates, used them for routine authentication over a semester-long course, and managed credentials across multiple devices. Our results show that initial setup and setting up credentials on a second device are difficult, while routine authentication is easy once configured. Nevertheless, perceived usability remained low, and alarmingly, only \pct{4} of participants fully understood mTLS security implications and key management. Our findings demonstrate that usability challenges shift across the credential lifecycle depending on system architecture. We conclude by offering design recommendations for future key management systems to better support users across the complete credential lifecycle.

\end{abstract}

%% file: sections/1.intro.tex
\section{Introduction}
\label{sec:intro}

Public key authentication methods offer a promising alternative to passwords~\cite{9152694}. However, such methods are only usable when they support users through the entire credential lifecycle, far beyond the initial login step~\cite{temoshok2024digital}. Historically, research on this lifecycle has centered on end-to-end encrypted or signed email, a domain that forces users to actively share keys and consistently highlights the difficulty of key management (see \S\ref{sec:rw}). Consequently, it remains unclear whether the usability issues come from public-key systems as a whole or are merely artifacts of the email signing tools. Resolving this ambiguity requires examining key management through a different lens.

We use Mutual Transport Layer Security (mTLS) as a case study. In standard web TLS, the browser verifies the server's identity. In contrast, mTLS also requires the server to verify the client's identity using a client certificate and proof that the client holds the matching private key~\cite{parsovs2013practical}. From a user perspective, mTLS requires navigating the complete credential lifecycle (see \S\ref{sec:background}), making it an ideal model for exploration. Therefore, we ask the following research questions:

\begin{enumerate}
  \item[\textbf{RQ1:}] What key management challenges do users encounter with mTLS?

  \item[\textbf{RQ2:}] Do these challenges shift across the credential use life cycle, i.e., initial setup, routine authentication, and cross-device migration?

\end{enumerate}

To answer these questions, we conducted a three-phase, semester-long study with 46 senior undergraduate and graduate students enrolled in an applied cryptography course. All participants had completed an introductory cybersecurity course and had basic knowledge of public-key cryptography. We refer to them as \emph{technical users}: they had relevant technical training but were not expert developers or PKI specialists and had no prior experience configuring mTLS. Given this training, we expect them to encounter fewer difficulties than everyday users without a technical background, although our study did not test this comparison or establish how everyday users would perform.

We asked participants to generate a key pair and certificate signing request (CSR), obtain a certificate from the course certificate authority (CA), register it with a browser, and authenticate to a course grading server. Second, they used mTLS to authenticate to this server over the semester for their project. Later in the semester, they authenticated from another device. We collected CA and authentication logs and analyzed these alongside participants' qualitative and quantitative written responses. In summary, we found the following:

\begin{enumerate}

  \item The CA rejected 62\% of the 416 CSR submissions. Among participants who failed their initial attempt ($n=38$, \pct{38}), it took an average of 4.5 submissions to finally have a CSR accepted. Even after successfully obtaining a certificate, participants struggled with the registration process due to unclear certificate-store locations, difficulty linking certificates to private keys, inconsistent browser behavior, and unhelpful error messages.

  \item Participants considered routine authentication easy after setup but still rated the overall mTLS experience poorly at the end of the semester.
  
  \item When setting up another device, participants had no built-in synchronization option, so they relied on the manual transfer of credentials or repeated enrollment, often choosing the option they perceived as easiest over security. Half of those who described their storage practices also kept credentials in local folders or private Git repositories.

  \item Even after a semester of use, only a few participants ($n=4$, \pct{4}) demonstrated strong understanding across all the security and credential-management issues examined.

\end{enumerate}

Drawing on these results, we show that key management challenges can arise at different stages of the credential lifecycle depending on the system architecture. As summarized in \S\ref{sec:discussion} (Table~\ref{tab:credential_lifecycle}), phases challenging in end-to-end encrypted email systems, such as sharing and using credentials, are straightforward in mTLS, whereas typically effortless stages, like generating and registering credentials, present the primary obstacles. We therefore offer design recommendations for future key management systems.

\paragraph{Paper Structure.}

The rest of the paper is organized as follows. \S\ref{sec:background} explains mTLS step by step from the user's perspective, followed by related work on authentication, secure email, client certificates, credential lifecycles, and security understanding (\S\ref{sec:rw}). We then detail our methodology (\S\ref{sec:methodology}), present findings (\S\ref{sec:results-setup}---\ref{sec:user-understanding}), and discuss limitations (\S\ref{sec:limitations}) before drawing broader lessons for key management (\S\ref{sec:discussion}) and conclude in (\S\ref{sec:conclusion}).


%% file: sections/2._background_and_related.tex
\section{Background}
\label{sec:background}

Transport Layer Security (TLS) underpins HTTPS, which secures most web requests and is used by the vast majority of websites~\cite{franken_2026_18246555}. Standard TLS authenticates the server, whereas mutual TLS (mTLS) also authenticates the client using a certificate and its corresponding private key~\cite{rescorla2018transport}. For browser users, this introduces a credential lifecycle consisting of three phases, which we detail below.

\subsubsection*{Enrollment and registration.}
Before first use, the user or client software generates a public--private key pair, retains control of the private key, and creates a CSR. Under PKCS\#10, a CSR contains a subject name, public key, optional attributes, and a signature created with the corresponding private key, thereby proving possession of that key~\cite{nystrom2000pkcs}. The user submits the CSR to a certificate authority (CA) through an enrollment service. The CA authenticates or authorizes the requester under its enrollment policy, verifies the CSR signature and policy requirements, and issues an X.509 certificate binding the approved identity to the public key~\cite{cooper2008internet,nystrom2000pkcs}. The user then imports the issued certificate and ensures that it is associated with the corresponding private key in the credential store used by the browser. PKCS\#12 can combine private keys, certificates, and related identity information in a protected container~\cite{10.17487/RFC7292}. A public certificate alone is insufficient because authentication requires the corresponding private key. Credential stores and import procedures vary across browsers and operating systems, and deployment studies document differences in browser behavior, certificate selection, and management interfaces~\cite{parsovs2013practical,10.1145/2974927.2974938}.

\subsubsection*{Routine authentication.}
When the user visits a protected site, the browser or operating system locates an appropriate installed client certificate and may prompt the user to select one when multiple certificates are available. The browser then uses the corresponding private key to authenticate the user without transmitting the private key itself~\cite{rescorla2018transport}. The server verifies that the certificate is valid, trusted, and authorized for the requested service before mapping its identity to a local account~\cite{cooper2008internet}. If these checks succeed, the application grants access. From the user’s perspective, routine authentication generally requires only visiting the site and, when prompted, selecting a certificate; the remaining steps occur automatically with little or no visible interaction.

\subsubsection*{Credential management.}
After enrollment, the credential must be managed throughout its lifecycle. To authenticate from another device, the user must either transfer the existing certificate and private key through a protected export-and-import process or generate a new key pair and enroll a separate certificate. PKCS\#12 supports transferring private keys and certificates between systems and applications~\cite{10.17487/RFC7292}. Certificates expire and may require renewal. If a certificate is lost while its private key remains available, the user may be able to obtain a replacement under the CA's policy. Loss of the private key requires a new credential, while suspected compromise additionally requires revocation and replacement~\cite{cooper2008internet}.

%% file: sections/2._rw.tex
\section{Related Work}
\label{sec:rw}

\subsection{Authentication Alternatives}

Bonneau et al.~\cite{6234436} and Zimmermann et al.~\cite{zimmermann12018quest} show that no universal password replacement exists; authentication methods must instead match their operational context and target users. 
Evaluating this fit requires looking beyond initial adoption to longitudinal usability. Reynolds et al.~\cite{reynolds2018tale} noted discrepancies between YubiKey setup and routine use, whereas Marky et al.~\cite{marky2022nah} demonstrated how security perceptions, friction, and effort govern multi-factor authentication acceptance. These lifecycle dynamics position FIDO2 and passkeys as the modern baseline for browser-based mTLS.

Nonetheless, research on FIDO2 adoption, workplace integration, and smartphone authentication reveals that successful setup rarely ensures sustained usability~\cite{9152694,farke2020you,3563572.3563576}. While passkeys leverage platform managers for device-bound or synchronized credentials to minimize user effort, they introduce fresh complexities regarding recovery, regulation, and organizational culture~\cite{3698900.3699304}. Further compounding these challenges are biometric misconceptions~\cite{lassak2021s}, challenges during continued use~\cite{308911}, inconsistent website implementations~\cite{bhardwaj2026state,jannettstate}, and interpersonal threats~\cite{daffalla2025framework}. Taken together, this literature suggests that platform support shifts rather than eliminates credential-management burdens across enrollment, routine authentication, and additional-device provisioning.

\subsection{Secure Email Systems}

Historical research into secure email systems provides extensive evidence regarding the usability challenges of user-facing public key management \cite{ruoti2019johnny}. Whitten and Tygar~\cite{whitten1999johnny} found that only a minority of participants successfully completed the full PGP workflow. A later evaluation of a modern PGP client showed that tighter webmail integration did not remove central usability problems \cite{ruoti2015johnny}. Paired studies of newer tools continued to identify difficulties with recipient discovery, trust decisions, protection selection, and regular adoption \cite{ruoti2016we, ruoti2019usability, bai2016inconvenient, borradaile2021motivated}.

Alternative cryptographic approaches like S/MIME address parts of this lifecycle by integrating encryption into mail clients and binding keys to certificates. Automated key collection and continuity cues help, but users still struggle with new identities and the meaning of encryption \cite{garfinkel2005johnny}. Other studies document difficulties obtaining certificates and configuring clients~\cite{fry2012not, banday2014s}. Measurements by Stransky et al.~\cite{stransky202227} further show that technical availability rarely translates into actual adoption. 

Integrated designs show that guided initiation, contextual instruction, and visible protection boundaries can improve practical usability, although automation can still create confusion about when protection is active \cite{ruoti2013confused, ruoti2016private}. Ruoti et al.~\cite{ruoti2018comparative} later held the interface constant while comparing password-based encryption, public key directories, and identity-based encryption (IBE). They found that automation eased setup and exchange, while it sometimes obscured the underlying security model and led to user misconceptions. Together the PGP, S/MIME, and IBE shift rather than eliminate lifecycle challenges. This pattern motivates examining browser-based mTLS, which retains public-key credential management without email-specific decisions.

\subsection{Client-Side Certificates}

Empirical evidence regarding client side certificates remains comparatively sparse. Manual enrollment took eight highly educated users an average of 140 minutes before Balfanz et al.~\cite{10.1109/MSP.2004.71} automated the process. Parsovs \cite{parsovs2013practical} found browser, server, selection, privacy, and portability problems across eighty-seven Estonian services, while Uda and Shikida \cite{10.1145/2974927.2974938} described a broad university deployment. Neither examined the complete user lifecycle. Subsequent architectural designs have attempted to ease individual stages. Dietz et al.~\cite{dietz2012origin} removed certificate selection but exposed account choice and portability limits. Conners et al.~\cite{conners2022let} automated issuance, management, and recovery while calling for longitudinal evaluation. At scale, Dong et al.~\cite{10.1145/3646547.3688415} found untrusted, shared, expired, and privacy-sensitive certificates in over one billion mTLS connections, while other measurements reveal tracking risks~\cite{foppe2018exploiting, 10.1145/3232755.3232763, 9511513}. Server certificate research also shows configuration difficulty and the value of automation~\cite{3241189.3241293, Tiefenau19CertbotUsable}, but does not capture how people create, register, use, and move browser client credentials.

\subsection{Lifecycle Management}

A critical gap in existing credential research is the phase following initial adoption. Secure email emphasizes initial trust and exchange, while certificate portability can conflict with protection against key extraction~\cite{dietz2012origin}. Passkey synchronization reduces transfer friction but shifts trust to providers and retains recovery needs~\cite{3698900.3699304, jannettstate}. Studies find inconsistent backup guidance~\cite{gerlitz2023adventures}, storage strategies with availability, confidentiality, and comprehension problems \cite{holtervennhoff2024mixed}, and backups that restore dependence on passwords, email, or telephone service \cite{gilsenan2023security}. Smith et al.~\cite{smith2023if} instead evaluated setup, authentication, removal, replacement, and auditing together. Migration, backup, revocation, and replacement thus stand out as core usability tasks.

\subsection{User Understanding}

System convenience often conceals critical gaps in user understanding. Users simplify encryption to restricted access \cite{3291228.3291260}, hold incomplete HTTPS models \cite{8835228}, and struggle with certificate trust and validation errors even in information technology roles \cite{10.1145/3419472}. Misconceptions also surround WebAuthn biometrics and automated secure email \cite{lassak2021s, ruoti2018comparative}. These findings support evaluating comprehension directly, especially for loss, compromise, migration, revocation, and recovery scenarios that may remain invisible during routine authentication.

\subsection{Research Gap}

Synthesizing this literature reveals how different credential architectures redistribute administrative and cognitive burden. PGP foregrounds discovery and trust, S/MIME adds issuance and configuration, identity-based email automates discovery through a trusted service, and passkeys rely on platform managers. In contrast, browser mTLS binds an issued certificate to a local private key that the browser presents automatically. No prior study follows the same users through browser mTLS setup, routine authentication, and use on another device. The existing literature therefore cannot separate general public key burdens from those introduced by CAs, command-line tools, operating system stores, and browsers. This study examines all three lifecycle stages through system records and user accounts, treating mTLS as a case study for understanding lifecycle differences rather than as a replacement for passkeys or a priority over better-supported passwordless systems.

%% file: sections/3._participants.tex
\section{Methodology}
\label{sec:methodology}


\begin{table}
\caption{Overview of study methodology and data.}
\label{tab:study-stages}
\centering
\footnotesize
\setlength{\tabcolsep}{2pt}

\begin{tabularx}{\columnwidth}{@{}
>{\raggedright\arraybackslash}p{0.38\columnwidth}
>{\raggedright\arraybackslash}p{0.22\columnwidth}
>{\raggedright\arraybackslash}X @{}}
\toprule
\textbf{Activity} &
\textbf{Data source} &
\textbf{Data collected} \\
\midrule

Generated credentials, configured a browser, and authenticated.
(\S\ref{procedure:initial_setup}) &
CA and authentication-gateway logs &
CSR rejection counts and reasons, key sizes, issued certificates,
CSR and certificate timestamps, authentication failures, and
first-success timestamps. \\
& Initial written reports (semester start) &
Self-reported setup time, experience, ASQ, and SUS. \\

\addlinespace\midrule\addlinespace

Accessed the server via mTLS for five project submissions.
(\S\ref{procedure:routine_authentication}) &
Authentication-gateway logs &
Authentication attempts associated with routine project submissions. \\
& Semester-end written reports &
Routine-use experience, credential-storage practices, and final SUS. \\

\addlinespace\midrule\addlinespace

Authenticated on a second device via credential transfer or new
generation. (\S\ref{procedure:multi_device_task}) &
CA logs &
New certificate issuance and timestamps, authentication outcomes. \\
& Semester-end written reports &
Self-reported setup time, migration approach and rationale,
reported problems, experience, and ASQ. \\

\addlinespace\midrule\addlinespace

Answered security questions.
(\S\ref{procedure:participant_understanding}) &
Semester-end written reports &
Answers subsequently coded as correct, partially correct, or
incorrect for each scenario, and an overall score from 0--6. \\

\bottomrule
\end{tabularx}
\end{table}

\subsection{Overview}
\label{sec:overview}
 
We conducted an IRB-approved, semester-long study ($n=46$) examining  how technical users configure and use client-side mTLS authentication.  We analyzed CA and server logs alongside qualitative and quantitative  measures of participant experience. 

Our study is based on a project students completed as an assignment in an applied cryptography class cross-listed for both senior undergraduate and graduate students at a public university. Across two semesters, 46 of 82 enrolled students (56\%) consented to participate (Semester~1: 22/38=58\%, Semester~2: 24/44=55\%). We refer to consenting participants as P1--P46. As a sensitivity check for aggregating two semesters, we compare key descriptives from Cohort~1 to the aggregated dataset after adding Cohort~2 (Appendix~\ref{sec:supplementary-results}, Table~\ref{tab:per-cohort-outcomes}); because this does not materially change summary outcomes or qualitative conclusions, we report aggregated results instead of cohort-by-cohort analysis. Participants were mostly male (about 20\% female, 80\% male), consistent with our department's demographics. Full details of consent and ethical considerations are in Appendix~\S\ref{sec:ethics} (\nameref{sec:ethics}).

Briefly, the mTLS activity was a required course assignment, but participation in the research study was not. We did not disclose the research aims during the semester and did not collect research consent at that time. After final grades were submitted, students were contacted and invited to provide voluntary consent for us to use their written reflections and relevant log data for research purposes, in accordance with IRB requirements. Students could decline without penalty, and non-consenting students’ data were excluded from all analyses. The activity was primarily educational, and we prioritized its learning objectives over research needs; therefore, participants were not paid.

\subsection{Educational Context}
\label{sec:educational-context}

\subsubsection{Prior Coursework} Students entered the applied cryptography course having already completed an introductory cybersecurity course, so they were not novices to security concepts. The applied cryptography course provided more in-depth instruction, covering mathematical primitives, cryptographic constructions, and their use in protocols like mTLS. Before the project, the course dedicated three class periods to authentication fundamentals, covering passwords, password-authenticated key exchange (PAKE), and multi-factor authentication, emphasizing common threat models and tradeoffs between passwords and possession-based authenticators \cite{6234436}. Students received roughly half a class period of lecture on certificate loss and revocation, covering TLS server-side certificates, key loss, and the general principles of server-side certificate verification. While this provided foundational knowledge of certificates, key management, and authentication, it did not address client-side certificate authentication in mTLS.

\subsubsection{Project Rationale} We chose mTLS as the course project to give students hands-on experience with key management. Key management is routinely underestimated as a deployment challenge, even by those building security 
systems~\cite{219400,ellison2000ten}, and this exposure prepares students for deploying cryptographic systems in their careers.

\subsubsection{Instruction Timeline} During the first part setup, students had not yet received any lecture instruction on mTLS. We verified prior exposure by asking participants whether they had previously used or configured mTLS or client certificates, and none reported prior hands-on experience with mTLS. Between the first and second setup, students attended a lecture that covered the TLS topics described above and included a brief, roughly five-minute introduction to mTLS, illustrating two-way authentication as an extension of TLS and nothing beyond that. The first project setup, therefore, represented students’ initial hands-on experience with mTLS, requiring them to apply their theoretical understanding of PKI and TLS to mutual authentication using client certificates.

\subsection{Server Setup}
\label{sec:server-setup}

We created a course CA to issue the client certificates used in the study. Students signed in through the university's Single Sign-On system and submitted a certificate request. The CA checked that the university ID in the request matched the signed-in student, preventing students from requesting certificates for other identities. The CSR had to include the participant identifier as the \texttt{UID}, the full name as the Common Name (CN), and institutional fields (Organization, Organizational Unit, Locality, State, Country). We configured the CA to require keys with at least 128-bit equivalent security (3072-bit RSA), enforcing a future-proof standard that exceeds current NIST minimums~\cite{barker2018recommendation,barker2018transitioning}. We also deployed a course grading server that checked intermediate assignment results. An NGINX reverse proxy acted as an authentication gateway in front of this server~\cite{NGINXSSL}. The gateway required students to present a certificate issued by the course CA before forwarding their requests to the grading server. Each successful or failed certificate check was logged.

\subsection{Study Procedure}
\label{sec:study-procedure}

\subsubsection{Initial Setup} 
\label{procedure:initial_setup}

We released the initial setup assignment before students had received any lecture content on TLS, and they were given one week to complete it independently as a take-home task on their own computers rather than in a controlled laboratory setting. Because the department requires personal laptops for coursework, students were expected to have a personal device, although university computing facilities were available. Although we did not record how many students used these facilities, the availability reduced, but did not eliminate, the need to use personal devices because the study later required participants to add a second device (\S\ref{procedure:multi_device_task}). We detail the ethics around personal device use in Appendix~\ref{sec:ethics}.

The assignment required students to configure browser-based mTLS to access a course grading server that automatically checked assignment outputs. Students generated a key pair, submitted a CSR, obtained a signed certificate, imported it into their browser, and authenticated to the server. We expected them to troubleshoot the setup and explore resources independently. The instructions did not specify the key requirement; students were expected to determine a compliant key size themselves. Instructor support was limited to cases where students had made an apparent but unsuccessful effort, reflecting real cryptosystem deployment where documentation may be incomplete and success often requires persistence and resourcefulness~\cite{3698900.3699303}. To support replication, our artifact package includes the exact participant instructions, an mTLS setup guide, and an error-to-fix guide covering common configuration problems (\S\nameref{sec:open_science}).

After 90 minutes of cumulative work time, students were encouraged to contact TAs for guidance via office hours, email, or the course messaging system. TAs provided hints or troubleshooting assistance but did not complete the steps for students. After three hours of cumulative work time, TAs were authorized to complete the setup on a student's behalf if needed, ensuring all students could access the course grading server for subsequent assignments. Students could request help at any point during the one-week assignment period. We established these thresholds to align with course-load expectations and ensure participants could complete concurrent coursework without the project consuming excessive time. 

Only a few students ($n=9$, \pct{9}) asked TAs for help, and none required a TA to complete the task on their behalf. All of the students whose work we evaluated finished the setup themselves. Notably, we were unable to verify how long these nine students spent before seeking help, or whether they actually reached the intended 90-minute threshold. However, they demonstrated progress when they sought help. We also recognize that this assistance policy and the lack of time tracking may reduce research validity. Nevertheless, we prioritized preserving the assignment's value as an educational experience over research goals.

Participants documented their experiences through open-ended written reflections immediately after the initial setup (Study materials in \S  \nameref{sec:open_science}). The reflections were submitted as part of the course requirements. However, they were graded only for completion, not for content accuracy or depth of reasoning, ensuring that responses remained authentic and uninfluenced by performance incentives. These first reflections captured the steps participants followed, the time required, the challenges encountered, and the strategies used to troubleshoot issues. Participants described which tools they adopted or abandoned, what support resources they consulted, and which steps felt intuitive or confusing during the initial configuration process. To complement qualitative reflections, we administered the After-Scenario Questionnaire (ASQ)~\cite{10.1145/122672.122692} and the System Usability Scale (SUS)~\cite{brooke1996sus} to capture task-specific difficulty and the initial perceived usability of TLS client authentication, respectively.

\subsubsection{Routine Authentication}
\label{procedure:routine_authentication}

After completing the setup, participants confirmed they could log in to the course grading server and then used mTLS to authenticate as they completed five subsequent course projects. We logged all authentication attempts as described in \S\ref{sec:server-setup}. At the end of the semester, we collected  written self report about their experience, which also included a follow-up SUS to measure how usability perception changed after sustained use.

\subsubsection{Multi-Device Task}
\label{procedure:multi_device_task}

After submitting their last assignment, we asked participants to authenticate from a secondary device to test mTLS credential portability and recovery. Those who initially used a personal computer switched to a shared university machine, and vice versa—a requirement that reduced, though did not eliminate, personal-device use. As with the initial setup, students completed the task unsupervised and on their own time. They had access to scheduled university computing facilities and received limited TA support, consisting of hints provided only after an independent attempt.
Participants could either transfer their existing certificate and private key from their original device or repeat the setup process to generate new credentials for the second device. We did not prescribe a particular method; instead, we encouraged students to make their own choice based on their understanding of the system. The course grading server verified successful completion via a dedicated pass-off link accessible only after a valid TLS handshake from the new device. 

They then completed a second report, which was graded for completion rather than correctness. This report also included the ASQ, tracking participants’ perceived complexity of the second device setups. In the report, they answered the same prompts as in the initial report and explained whether they reused an existing certificate or generated a new one, along with their reasoning. This allowed us to observe their cross-device key management choices. Participants were also asked to provide a final, semester-long report, which included how they managed certificates and private keys during the study period.

\subsubsection{Participant Understanding} 
\label{procedure:participant_understanding}
While users may authenticate successfully without understanding PKI internals, accurate understanding matters when they must choose actions that affect security and availability. Accordingly, we measure understanding as whether participants can correctly state outcomes and select appropriate actions in exception scenarios. After the second device setup, we administered six scenario-based thought exercises. We began by asking participants to compare certificate-based and password-based authentication to explore their understanding of security trade-offs. Next, we prompted them to reason about the trade-offs between reusing and regenerating certificates across devices. Finally, we examined their understanding of credential compromise by asking how they would respond to certificate loss, key loss, and key theft, and how a CA should react in cases of suspected misuse.

\subsection{Data Analysis}
\label{sec:data-analysis}

\subsubsection{Server Log Analysis}

We excluded non-consenting participants and invalid PEM-formatted CSRs from server logs. Using Python's \texttt{cryptography} library~\cite{cryptography}, we created a canonical identifier for each submission from its CSR Common Name (CN) and participant ID, obtained from the \texttt{UID} or email local-part. For rejected CSRs, we flattened returned error lists and tallied distinct messages to identify common failure modes; we also extracted public-key sizes to compute key-size distributions for all submitted and accepted CSRs. For issued certificates, we mapped serial numbers to participants to measure multiple-certificate issuance (re-enrollment/rollover). Finally, we analyzed authentication logs to quantify setup difficulty (successes/failures per participant), focusing on failures before the first successful login, and computed first-attempt success rates and failure-to-success ratios to characterize troubleshooting iterations.

\subsubsection{Time analysis}

We measured time as the interval between a participant’s first failed authentication and first successful handshake. This does not include any work done before the first failure, includes any breaks between attempts, and cannot be computed for participants who succeeded on their first attempt, so they were excluded. We also collected self-reported setup times. We report both separately because they cover different parts of the process and are not directly comparable. We use geometric means (GM) to summarize the skewed values. Because participants ($n=9$, \pct{9}) received TA help after an unknown amount of independent work, these values should be seen only as rough context, not exact setup times.


\subsubsection{Quantitative Methods}
\label{sub:quantitative-methods}

\begin{table}
    \caption{Variables and roles in the exploratory analyses.}
    \label{tab:analysis_variables}
    \centering
    \small
    \setlength{\tabcolsep}{3pt}
    \begin{tabular}{@{}p{0.21\columnwidth}p{0.25\columnwidth}p{0.48\columnwidth}@{}}
    \toprule
    \textbf{Variable group} &
    \textbf{Analytical role} &
    \textbf{Measures} \\
    \midrule

    Perceived usability &
    Outcome (dependent variables) &
    Initial SUS, initial ASQ, end-of-semester SUS, and second-device ASQ. \\

    Setup and migration &
    Explanatory (independent variables) &
    CSR rejection count, browser or OS failure, setup time, migration approach, and number of certificates. \\

    Comprehension &
    Explanatory (independent variables) &
    Scenario-specific understanding categories and overall understanding score. \\

    \bottomrule
    \end{tabular}
\end{table}

We averaged the three ASQ items into a composite score and used a Friedman test~\cite{friedman1937use} to evaluate differences across ASQ dimensions. We scored SUS on its standard 0--100 scale~\cite{brooke1996sus} using established benchmarks~\cite{bangor2008empirical}. To address missing data, we calculated scores using only the available responses. Specifically, two participants (P12, P22) skipped the third ASQ item. we computed their composite from the two available items and excluded them only from the Friedman analysis. One participant (P41) did not complete the end-of-semester SUS and was excluded from that analysis. However, their qualitative responses were retained, as including participants with partial data is an established practice in mixed-methods research~\cite{creswell2007designing}.

We examined exploratory, correlational associations among setup, migration, comprehension, and usability, with variable roles defined in Table~\ref{tab:analysis_variables}.  Usability measures were generally treated as outcome variables, while setup, migration, and comprehension measures were generally treated as explanatory variables. These roles were analysis-specific. All variables were observed rather than experimentally manipulated; therefore, the results do not support causal inference.

Analyses used $n=46$ unless reduced by missing data and assumed independent between-participant observations. Test choice followed variable type and design: Spearman correlations for monotonic associations involving count or ordinal measures; Welch's $t$-tests~\cite{welch1947generalization} for two-group continuous outcomes (allowing unequal variances and assuming approximate normality and no severe outliers); Pearson's $\chi^2$ tests for categorical associations with adequate expected counts; Kruskal--Wallis tests~\cite{kruskal1952use} for nonparametric comparisons across comprehension levels; and Friedman tests for repeated within-participant ASQ ratings. We applied Benjamini--Hochberg FDR correction~\cite{benjamini1995controlling} at $q=0.05$ across 22 exploratory association tests, treating the ASQ analysis separately. Significant Kruskal--Wallis tests were followed by Holm--Bonferroni-corrected~\cite{holm1979simple} Mann--Whitney $U$ tests~\cite{mann1947test}. Table~\ref{tab:consolidated-stats} in Appendix~\ref{sec:supplementary-results} reports raw $p$-values, corrected $q$-values, and effect sizes.

\subsubsection{Qualitative Analysis}
We drew on prior experience analyzing qualitative data and publishing the resulting findings. Two of the authors used open coding and the constant comparative method~\cite{glaser1965constant, holton2007coding} to iteratively develop and refine the codebook. Rather than coding independently and calculating inter-rater reliability afterward, we coded all responses collaboratively, resolving discrepancies in real time to reach full consensus (Appendix~\ref{sec:open_science} details how to access the resulting codebook). To evaluate participants' security understanding, we then classified each response into three categories: \emph{Correct} for complete and accurate answer; \emph{Partially Correct} if it contained identifiable misconceptions alongside correct answer~\cite{smith1994misconceptions, kaczmarczyk2010identifying}; and \emph{Incorrect} if it revealed a fundamental misunderstanding of security implications.


%% file: sections/4._setup_m_r.tex
\section{Findings: Initial Setup}
\label{sec:results-setup}

From participants’ initial ASQ responses, the experience was quite neutral-to-negative (median $ASQ =3.67/7$). While participants rated ease ($\tilde{x}=4.0$) slightly above time and support ($\tilde{x}=3.0$), a Friedman test did not detect a significant difference among the three ASQ item ratings ($\chi^2(2)=3.29$, $p=0.19$), reflecting a generally subpar experience.Also, initial SUS responses (SUS $\mu=50$, grade ``F'') showed that perceived usability fell in the not acceptable range, with only \pct{5} rating it \textit{Acceptable} (SUS $\geq 70$). Interestingly, combining authentication-log-derived setup times with initial SUS responses, we found no association between setup time and perceived usability ($\rho=-0.008$, $p=.959$, $q=.988$; negligible effect). When looking at those completion times, self-reported estimates ($GM=2.2$ hours) were noticeably higher than the log-derived estimates ($GM=1.2$ hours). While we did not reconcile these two measurements at the individual level, participants reported completing the task intermittently.
 
\longsay{\dots took me maybe about 3 hours spread out in my day.}{P29}

To understand the factors associated with these poor usability scores, we detail the specific stage-level problems participants encountered during their initial mTLS setup below

\subsection{Key and CSR Generation}
\label{subsection:Key Pair Generation}

Here, we examine how key and CSR requirements created difficulties during the initial setup phase, before participants reached browser configuration.

Our CA logs recorded 416 CSR submissions, of which 258 (62\%) were rejected. The logs also showed that only a few participants ($n=8$, \pct{8}) produced an accepted CSR on their first submission, while the remaining participants ($n=38$, \pct{38}) averaged 4.5 attempts before producing an accepted CSR. These first-submission results characterize participants' initial path to an accepted CSR. In contrast, Figure~\ref{fig:csr-rejection-cdf} in Appendix~\ref{sec:supplementary-results} summarizes all rejected CSR submissions recorded across the full study period, rather than only those preceding a participant's first accepted CSR. 

Across the full study period, the CA logs showed that most participants ($n=43$, \pct{43}) experienced at least one rejection, and the median was 4 rejected submissions per participant. Accordingly, some participants whose first submission was accepted nevertheless experienced at least one rejection later in the study. Overall, these results demonstrate that CSR generation commonly involved repeated submission and correction.

\begin{figure}
    \centering
    \includegraphics[width=0.95\linewidth]{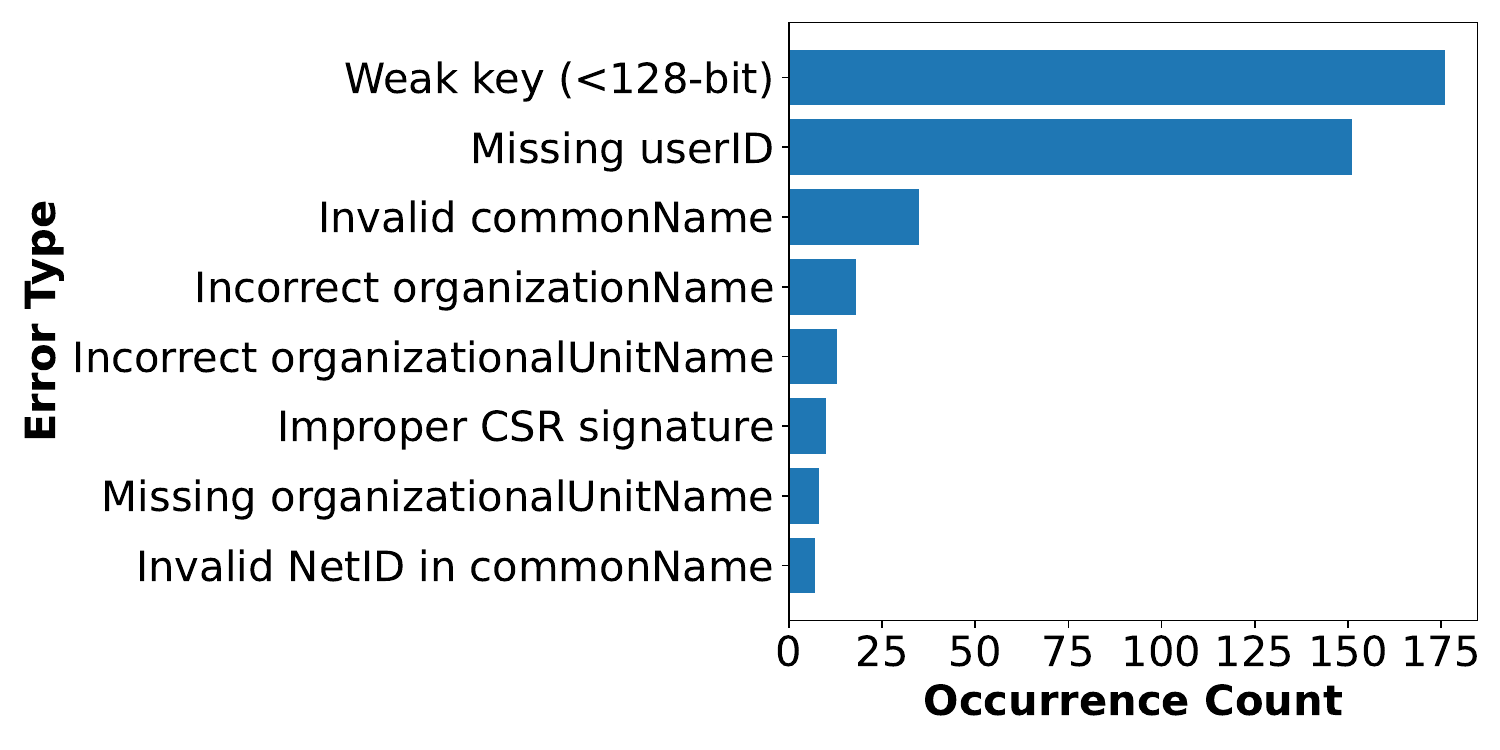}
    \caption{CSR rejection reasons ($\geq$5 occurrences)}
    \Description{A bar chart showing CSR rejection reasons that occurred at least five times. Weak cryptographic keys and missing userID attributes are the most frequent rejection reasons.}
    \label{fig:csr-errors}
\end{figure}

From the initial written reports, we found that nearly half ($n=22$, \pct{22}) described key generation as the easiest mTLS step. In contrast, some participants ($n=10$) reported having their CSRs rejected for unsupported key sizes. When we look at the CA logs to understand the prevalence of this problem, we found that key size was a frequent source of rejection during CSR creation (Figure ~\ref{fig:csr-errors}). We also found that of the 416 submitted CSRs, 131 (32\%) failed the course CA's 3072-bit key requirement (Figure~\ref{fig:keysize-initial} in Appendix~\ref{sec:supplementary-results}); 2048-bit keys alone accounted for 109 rejections (26\%). The same logs showed that 199 submissions (48\%) used 4096-bit keys. 

Interestingly, the initial written reports showed that all participants reported using OpenSSL, and the documented OpenSSL default was 2048-bit RSA~\cite{openssl_genrsa}. While trying to interpret this high key rejection rate as a result of the tool default, however, our CA logs do not establish whether that default caused participants to select 2048-bit keys. However, one participant's written reflection illustrates the difficulty of interpreting the key-size requirement:

\longsay{I didn’t really know why I was getting an error \dots Only after reading more carefully (at least 128-bit security) did I try to go higher and I switched to RSA 4096.}{P35}

Our CA-log analysis also found that 81\% of successfully accepted CSRs used 4096-bit keys. When we compared the key-size distributions between initial attempts and accepted requests (Figure~\ref{fig:keysize-shift} in Appendix~\ref{sec:supplementary-results}), we observed a clear shift: 4096-bit keys were much more prevalent in the final accepted pool. While this pattern suggests that participants adjusted their key size after experiencing a rejection, we cannot definitively isolate the effect of error feedback. Because the accepted-CSR dataset combines participants who retried after an error with those who succeeded on their very first attempt, we cannot confirm whether the shift was driven entirely by error correction or simply by initial high-bit choices.

This raises the question of how much the course CA's comparatively strict policy contributed to the observed rejection rate. While commercial CAs may override malformed metadata~\cite{10.1145/2974927.2974938}, they cannot cryptographically strengthen a weak key supplied by a user. Nevertheless, the number of rejected submissions depends on the CA's acceptance threshold. Recalculating the CA-log data under a hypothetical 2048-bit minimum reduced the number of key-size rejections from 131 to 22 of 416 submissions (5\%). Thus, the observed rejection rate reflects an interaction between the course CA's policy and tools that did not clearly communicate or help users satisfy that policy. Lowering the threshold would reduce key-size rejections, but it would not address the browser and operating-system problems encountered later in the workflow (\S\ref{sec:mtls-browser-os}).

Also from the written responses, we see that most participants ($n=35$, \pct{35}) reported failed or malformed CSR creation attempts. The most frequent issue was the default OpenSSL command-line method ($n=10$). Some participants tried alternatives such as Puttygen or web-based CSR generators ($n=4$) but reported downstream compatibility problems before abandoning them. Others also struggled to customize attributes—producing malformed requests ($n=12$), experimenting with the UID field ($n=5$), or attempting to use unrecognized object identifiers ($n=2$). Individual missteps also included editing the OpenSSL config or misinterpreting standard X.509 fields, both of which ended in failure. Overall, CSR creation challenged not only participants' technical fluency but also their understanding of certificate standards and tooling.

Beyond self-reports, our CA logs contained 17 distinct rejection-error types and 436 error messages across the 258 rejected submissions, averaging 1.7 messages per rejected submission. Weak cryptographic keys and missing \texttt{userID} attributes were the most frequent recorded rejection reasons (Figure~\ref{fig:csr-errors}). Together, the written responses and CA logs indicate that CSR creation required participants to satisfy precise syntactic and policy requirements that were not adequately supported by their tools.

Surprisingly, CSR rejection counts were not significantly associated with initial SUS scores ($\rho=-.204$, $p=.173$, $q_{\mathrm{BH}}=.517$; small effect) or initial ASQ scores ($\rho=.198$, $p=.188$, $q_{\mathrm{BH}}=.517$; small effect). Thus, we look at the participant experience after they complete key and CSR generation in the next subsection.


\subsection{Registering the Certificate with the Browser}
\label{sec:mtls-browser-os}

Even after successfully obtaining a signed certificate, we found from written reports that most participants ($n=41$, \pct{41}) had Browser/OS UI problems. Nearly half of them ($n=18$) explicitly called this the hardest phase of the workflow.

\longsay{The hardest step by far was registering the certificate and key to my computer.}{P6}

Combining the browser-failure indicator derived from initial written reports with initial SUS responses, participants reporting browser failures had lower mean SUS scores (SUS: $48.08$ vs.\ $63.50$). Although this difference was not statistically significant ($t = -2.406$, $p = .058$, $q = .317$, $d = -1.021$), the large effect size suggests a potential association between browser level obstacles and lower perceived usability. Below, we detail these specific problems from the written reports. 

\subsubsection{Store discovery and placement, ($n=28$, \pct{28})} \label{sec:store-discovery}

Participants encountered multiple, partially overlapping import interfaces, including OS-level stores such as Windows Certificate Manager, OS keychains such as macOS Keychain Access, and browser-specific certificate managers. Participants cited uncertainty about which credential store the browser consulted as a reason for adopting trial-and-error strategies and imported their artifacts into multiple locations. P16 described this coping strategy:

\longsay{I was largely just mad \ldots\ so I just imported the cert into 5--6 different cert locations on my computer.}{P16}

Their response also identified terminology and organization mismatches. Participants had difficulty distinguishing website certificates from personal certificates and locating the appropriate browser-specific category. In other cases, a certificate appeared in an OS-level interface but remained absent from the browser's certificate manager. These inconsistencies required additional trial and error and left participants uncertain whether the import had succeeded or their setup remained incomplete.







\begin{mytakeaway} 
    Generating valid mTLS artifacts is only half the battle; users frequently fail simply when UIs fail to communicate \emph{where} to place certificates or \emph{how} to verify them.
    \end{mytakeaway}

\subsubsection{Key--certificate association and the ``hidden'' PKCS\#12 requirement ($n=35$, \pct{35})}
\label{sec:pkcs12}
 
Participants repeatedly reported that importing only the signed certificate was insufficient for authentication. In several cases, artifacts appeared ``installed'' within the UI but were never offered during the TLS handshake. Through external research and trial-and-error, users eventually discovered they had to bundle the certificate and private key into a single container format (commonly PKCS\#12: \texttt{.p12}/\texttt{.pfx}).

\longsay{I thought all I needed to do was import my CSR into my keychain access System certificates. However, this was not working after lots of trial-and-error also known as doing-the-same-thing-and-hoping-for-different-results. Finally, a site mentioned that I would need to generate a PFX with my private key.}{P29}

\begin{mytakeaway}
OS and browser UI often accept standalone certificates without warning users that authentication requires a bundled key-certificate container to establish proof-of-possession.
    \end{mytakeaway}

\subsubsection{Activation and selection behavior varies by browser and state ($n=15$, \pct{15})}
\label{sec:activation-selection}

Even after creating and importing a correct bundle, participants struggled to get the browser to actually offer the certificate during the TLS handshake. Participants frequently encountered browser-state quirks, relying on private/incognito windows or clearing the SSL state and caches, and adjusting browser settings. P31 described one successful workaround:

\longsay{I then tried by entering an incognito tab and reloading. This time, Chrome prompted me to choose a certificate \ldots}{P31}


Participants also reported cross-browser inconsistencies, where the exact same artifacts worked in one browser but failed in another.

\longsay{I still couldn't access the website through Edge. However, I was able to successfully reach it through Google Chrome.}{P41}

In some cases, participants only succeeded after restarting their browser or computer, which made it difficult to determine whether earlier failures were caused by incorrect configuration or transient system state.



\begin{mytakeaway}
    Valid mTLS setups routinely fail when browser state, stale caches, and inconsistent prompting policies masquerade as setup failures, leaving users unable to diagnose the true issue.

    \end{mytakeaway}

\subsubsection{Feedback and recovery are dominated by opaque errors and compatibility traps ($n=24$, \pct{24})}
\label{sec:errors-recovery}

Participants repeatedly highlighted that Browser/OS UIs provided weak diagnostic feedback when an installation failed. P5 directly connected this missing feedback to the difficulty of diagnosing setup problems:

\longsay{Registering the signed certificate and private key with the browser was the most difficult due to the lack of useful error feedback, which made it very difficult to diagnose issues with the setup.}{P5}

A recurring example was a PKCS\#12 import failure on macOS despite users providing a known-correct password. Participants discovered this required platform- and version-specific workarounds, such as appending OpenSSL legacy compatibility flags.

\longsay{The legacy flag was added later \ldots\ it was not accepting the password I knew to be correct.}{P29}

Other participants encountered more severe compatibility problems. Some resorted to downgrading OpenSSL because its newer PKCS\#12 defaults were incompatible with macOS graphical tools, while others used command-line recovery paths when the graphical interfaces provided no way forward.





\begin{mytakeaway} Unclear error messages left users uncertain about whether setup problems involved the OS, web browser, or file format.

\end{mytakeaway}

 


\subsection{Resource Use and Coping Strategies}
\label{Tools and Resources}

The initial written reports showed that participants sought external help, most frequently from community forums ($n=25$), vendor or tutorial documentation ($n=23$), official OpenSSL documentation ($n=16$), and AI assistants such as ChatGPT ($n=16$). Resource usefulness depended largely on specificity. Targeted resources helped participants include the UID attribute, create a PKCS\#12 bundle, import it into a browser, and apply the macOS legacy compatibility flag. To resolve the unfamiliar PKCS\#12 requirement, participants used manual pages to understand why the certificate and private key had to be bundled, then obtained the necessary commands from Stack Overflow or task-specific tutorials. When guidance matched the problem, it could substantially reduce effort:

\longsay{I ended up relying on ChatGPT, it streamlined this incredibly well. It would have taken me 10x + more time if I didn't have that resource (or TA help).}{P24}

Conversely, participants also encountered outdated or incomplete guidance:

\longsay{Stack Overflow was the most helpful in multiple stages. Most other sources were outdated or ended abruptly, especially the ones recommending Chrome imports via settings that no longer exist.}{P2}

Overall, broad documentation and undirected searches often resulted in trial and error, whereas task and platform specific guidance resulted in resolving concrete problems. Successful coping therefore involved decomposing the workflow into specific subproblems and locating targeted instructions.

\begin{mytakeaway}
We see that even highly technical participants found the first-time mTLS setup challenging. They struggled with complex tools, inconsistent OS workflows, opaque browser interfaces, and poor feedback.

\end{mytakeaway}

%% file: sections/5._usage_m_r.tex
\section{Findings: Routine Authentication}
\label{sec:phase_2}

\begin{figure}
  \centering
  \includegraphics[width=0.50\linewidth]{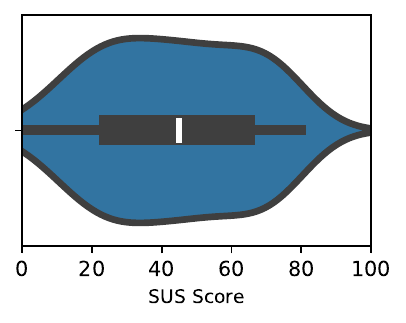}
  \Description{A box plot showing the distribution of System Usability Scale scores. The scores are concentrated between 26 and 63, indicating generally poor usability.} 
  \caption{Distribution of end-of-semester SUS scores.}
  \label{fig:sus-scores}
\end{figure}

The clearest positive finding concerned routine authentication after enrollment. In the end-of-semester report, which was collected after the multi-device task, participants were asked specifically about routine authentication. Among those who commented on routine use ($n=23$), participants generally described authentication as easy once setup was complete. P11 explained:

\longsay{The easiest thing about using TLS client authentication was the seamless access after the initial login.}{P11}

Despite these positive experiences with routine use, we also noted a stark contrast between setup difficulty and daily ease ($n=30$). Participants expressed strong frustration with the setup process while acknowledging the smooth experience that followed: 

\longsay{Setting up TLS authentication was hell. I hated every moment of it. Using it, however, was a cakewalk\dots}{P16}


Notably, at the end of the semester, students perceived mTLS as less usable than they had initially, despite daily use being relatively straightforward (SUS $\mu$ = 44, \textit{Not acceptable, grade F}, Figure~\ref{fig:sus-scores}). We found that only 15\% ($n=7$) rated the experience as \textit{acceptable} (SUS~$\geq$~70).

\begin{mytakeaway}
Participants perceived routine use as easy over the semester, yet final usability ratings remained unacceptable.
\end{mytakeaway}

\paragraph{Credential Management Practices}

In the end of semseter report, we asked how participants stored their signed certificate and private key over the semester, particularly before configuring a second device. Among those who described their practices ($n = 34$), half ($n = 17$) followed good security hygiene as defined by NIST~\cite{barker2020nist}, using integrated solutions such as macOS keychain ($n = 11$) or the browser’s certificate store ($n = 6$). The other half saved credentials in a local folder on their device ($n = 14$) or in private Git repositories ($n = 3$). While some participants recognized the risks of these methods, they made trade-offs for convenience or due to a perceived lack of urgency.


\longsay{I knew that it would be a pain if I ever lost the certificate or private key, so I just had all files \dots saved to GitHub repository. I'm aware that it's bad practice to push keys \dots to GitHub repositories but the repository is private and I am the only collaborator.}{P28} 

To assess the actual risk of relying on private repositories, we committed a dummy private key to a private GitHub repository. Instantly, we received an automated alert from GitGuardian~\cite{GitGuardian}, confirming that the key had been detected despite the repository being private. This exposes two concrete threats that participants did not consider: first, third-party scanning 
services already have read access to private repository contents, meaning the key is visible beyond the account owner; second, if the GitHub account itself is compromised---through credential theft, phishing, or session hijacking---the private key is immediately exposed alongside it, collapsing the security of the mTLS credential entirely. The perceived security of private repositories is therefore illusory. If even technical users make such mistakes, the risks will likely be amplified when mTLS is deployed to general users with less security awareness.



%% file: sections/6._setup2_m_r.tex
\section{Findings: Multi-Device Management}
\label{sec:phase_3}

Setting up a second device was considerably easier (median ASQ $=5.67/7$) and faster (self-reported $GM=0.48\mathrm{h}$) than the initial setup. Because synchronizing an existing credential did not require issuing a new certificate, CA logs could not fully identify participants' migration approaches. We therefore classified these approaches using their second written reports.

\subsubsection*{Migration Approach and Motivation}

Participants' second reports revealed a nearly even split: \pct{24} ($n=24$) synchronized an existing credential, while \pct{22} ($n=22$) generated a new one.

Among participants who synchronized an existing credential, the most frequently coded motivations were avoiding redundant setup ($n=10$) and perceiving file transfer as less labor-intensive than reenrollment ($n=6$). One participant synchronized after encountering errors while attempting to generate a new certificate, and another did so after experiencing difficulties with GUI forwarding. These findings indicate that synchronization was primarily selected for convenience or as a workaround for reenrollment problems. Among participants who generated a new credential, half ($n=11$) reported that repeating the now-familiar enrollment process seemed simpler while some ($n=5$) did not know that synchronizing an existing credential was possible:

\longsay{I didn’t know how to synchronize but I knew I could issue a new certificate relatively easily and quickly.}{P43}

Manual file-management barriers also influenced this decision. Forgotten export passwords ($n=2$) and restricted SCP access ($n=1$) led participants to generate new credentials instead. Only few participant ($n=2$) explicitly identified transferring the private key as a security concern.

Taken together, self report shows that more than three-quarters of participants (\pct{35}, $n=35$) relied on either an existing credential or procedural knowledge from the initial setup. Their written reports referenced existing \texttt{.p12}/\texttt{.pfx} files, saved commands, personal notes, and earlier project reports. This reuse provides qualitative context for the improved second-device ASQ ratings and completion time. 

Also, end-of-semester SUS scores did not differ significantly between participants who synchronized an existing credential and those who generated a new one (Welch's $t=-0.015$, $p=.988$, $q_{\mathrm{BH}}=.988$, $d=-0.004$; negligible effect). CA-log data further showed that \pct{37} of participants ($n=37$) used more than one certificate across the semester, with a mean of $3.5$ certificates per participant. Thus, the approach selected during the second-device task did not necessarily represent a consistent semester-long strategy: participants could synchronize an existing credential for one device while issuing additional certificates at other points.

\subsubsection*{Migration Pain Points}

Despite the reduced completion time and improved ASQ ratings, participants’ second written reports still described cross-device credential management as burdensome:

\longsay{The primary drawback is the increased overhead \dots for key management. Syncing certificates and keypair between multiple devices a user may interact with is a headache, especially when it needs to be done at scale and done in a secure manner.}{P27}

Among the 22 participants who generated a new certificate, half of them ($n=11$) reported problems creating, importing, or updating certificate files and keys. Thus, prior experience with the initial setup did not guarantee successful repetition of the procedure. Participants also reported difficulty recreating or packaging CSRs and verifying their correctness. Browser integration remained a recurring problem, with participants ($n=4$) specifically reporting difficulty locating or importing certificates on the second device.

Participants who synchronized an existing credential encountered and reported a different set of problems. Without a native browser export workflow, participants relied on raw file transfers and  encountered edge-case synchronization failures. A few reported X11 forwarding issues for browser access ($n = 1$), difficulties hosting temporary HTTP servers for file transfer ($n = 1$), or problems connecting to the pass-off server ($n = 1$). Certificate transfer and connectivity failures ($n = 5$), often involving missing configurations or multi-factor authentication requirements when using tools such as SCP, were also reported. 



\subsubsection*{Security Understanding}

Furthermore, participants showed mixed understanding of the security implications of the two certificate migration approaches: nearly two-thirds ($n=28$, \pct{28}) fully grasped the trade-offs, while the rest did not (Table~\ref{tab:understanding-overview}). This indicates that even technical users may have limited comprehension, suggesting that extending client authentication to general populations could increase security risks when multiple devices are used, potentially undermining the protective goals of authentication systems. 

Interestingly, a complete understanding of the migration trade-offs did not translate into a clear preference for one approach. Among the participants ($n=28$) with complete understanding, only about half ($n=15$) chose to synchronize their existing certificate. Consistent with this split, complete versus incomplete understanding was not significantly associated with certificate approach ($\chi^2(1)=0.534$, $p=.465$, $q=.731$, Cramér's $V=.109$; small effect). The same pattern extended to participants' usability ratings: end-of-semester SUS scores did not differ significantly across the three levels of understanding of the synchronization-versus-new-certificate trade-off (Kruskal-Wallis $H(2) = 2.134$, $p = .344$, $q_{\mathrm{BH}} = .582$, $\epsilon^2 = .003$; negligible effect).

\begin{mytakeaway}  In the absence of native synchronization features in browser/OS UIs, participants resorted to manual file transfers, exposing them to risks some did not fully understand and even those who understood still chose what was easiest.
\end{mytakeaway}

\subsubsection*{Migration Improvement}
\label{sec:sync-recommendations}

To address the key migration challenges identified above, we analyzed participants' feedback ($n=38$). We found that participants wanted improvements to the browser experience ($n=13$), clearer migration documentation ($n=8$), centralized and automated key management ($n=4$), more secure ways to transfer credentials ($n=4$), and better platform support, including headless and command-line workflows ($n=2$). Also, some participants ($n=7$) were satisfied with the current process and saw no need for changes.  These participant-identified priorities, together with the problems observed during initial setup (\S\ref{sec:results-setup}), inform the broader design implications for key-management systems discussed in \S\ref{sec:discussion}.

%% file: sections/7._understanding.tex
\section{Findings: Participants' Understanding}
\label{sec:user-understanding}

An analysis of the participants' written reports showed that, despite completing a full course in applied cryptography, they averaged only three correct answers out of six security reasoning questions (median = 3.0, mean = 3.15). Only a few participants ($n=4$, \pct{4}) achieved a perfect score; instead, the reports revealed lingering misconceptions in a substantial portion of the cohort (\pct{42}), as detailed in the subsections below.

\begin{table}
    \centering
    \small
    \setlength{\tabcolsep}{6pt}
    \caption{Understanding Across Security Questions}
    \label{tab:understanding-overview}
    \begin{tabular}{lrrr} 
    \toprule
    \textbf{Security Scenario} & \textbf{Correct} & \textbf{Partial} & \textbf{Incorrect} \\
    \midrule
    mTLS vs. Password & 32 (\pct{32}) & 10 (\pct{10}) & 4 (\pct{4}) \\
    \rowcolor{black!5} 
    Sync vs. New Cert & 28 (\pct{28}) & 9 (\pct{9}) & 9 (\pct{9}) \\
    Lost Cert (Key Intact) & 10 (\pct{10}) & 23 (\pct{23}) & 13 (\pct{13}) \\
    \rowcolor{black!5}
    Lost Private Key & 20 (\pct{20}) & 16 (\pct{16}) & 10 (\pct{10}) \\
    Stolen Private Key & 29 (\pct{29}) & 12 (\pct{12}) & 5 (\pct{5}) \\
    \rowcolor{black!5}
    Cert Detection and Revocation & 26 (\pct{26}) & 8 (\pct{8}) & 12 (\pct{12}) \\
    \bottomrule
    \end{tabular}

\end{table}

\subsubsection*{mTLS vs Password}

In contrast to passwords, which all participants understood intuitively, we observed two opposing misconceptions about mTLS among those with incomplete understanding of its advantages and limitations ($n=14$). Some participants viewed mTLS as essentially flawless, overlooking risks such as private key theft or certificate compromise:

\longsay{...Security wise, there are no downsides}{P31}

Conversely, participants focused on setup difficulty, believing mTLS offered no real security advantages. 

\longsay {I think that it is nice that it puts less burden on the user to remember passwords but it does not feel like it gives any more security benefits. The drawback is that it is much harder to set up the first time.}{P36}

Regardless of how accurately participants understood these trade-offs, their overall usability ratings remained similarly poor. End-of-semester SUS did not differ significantly across the three understanding levels (Kruskal--Wallis $H=2.201$, $p=.333$, $q=.582$, $\varepsilon^2=.005$; negligible effect).

\subsubsection*{Certificate Loss}

The scenario of losing a certificate while retaining the private key revealed the most severe comprehension gap in our study: only \pct{10} ($n=10$) demonstrated complete understanding of this critical distinction. We found that many participants conflated certificate loss with complete credential compromise, incorrectly assuming the entire identity was invalid. Despite the scenario explicitly stating that the key was safe, participants mistrusted this condition and worried about compromise:

\longsay{I would need to regenerate a new CSR using the private key from the lost certificate. I can submit that CSR to get a new certificate. The issue with this approach is that the key might be compromised, allowing an attacker to impersonate me}{P34}

This caution, while perhaps security-conscious, reflects confusion between actual and hypothetical threats. Even when explicitly told which component was compromised, participants struggled to reason reliably about the security implications.

\subsubsection*{Private Key Loss}

Twice as many participants ($n=20$, \pct{20}) recognized that losing a private key requires complete re-enrollment, compared to certificate loss with an intact key ($n=10$). Of the six scenarios, understanding of private-key loss showed the most noticeable variation across understanding levels (See Table~\ref{tab:consolidated-stats} in Appendix ~\ref{sec:supplementary-results}). Also, we found two critical misconceptions amongst those who did not fully understand the threat ($n=26$). First, some participants believed certificates could function without their corresponding private keys, a fundamental misunderstanding of public-key cryptography:

\longsay{If you still have a certificate, nothing. The private key is only used for creating the CSR, after which it is baked into the certificate}{P26} 

A separate misconception was noted regarding how to respond after losing a private key. For example:


\longsay{reverse engineering a csr and private key from a crt}{P42} 

\subsubsection*{Key Theft}

Key theft was the most intuitive scenario for participants: nearly two-thirds ($n=29$, \pct{29}) correctly recognized it as a severe event. However, others failed to distinguish passive loss from active compromise, treating an adversary-controlled private key with the same response as misplacing a file, revealing a gap in threat modeling.\longsay{If your private key was stolen, you would need to perform the same steps as if it was only lost.}{P11}

Even participants who recognized that revocation was necessary rarely knew how to carry it out. This gap between knowing what must happen and knowing how to act left participants uncertain about how to respond during a real security incident:

\longsay{You would need to prove that your private key was stolen or lost, which could be difficult \dots I am not entirely sure how to address this problem though.}{P18}

\subsubsection*{Detecting Certificate Theft}

When asked how a CA or course grading server could detect and respond to certificate theft, participants ($n=20$, \pct{20}) did not fully understand the problem. Several assumed that a stolen certificate would somehow look different in use:

\longsay{The CA could identify that it is stolen by comparing the websites that have been used previously by that certificate, and if there are any inconsistencies, we can potentially know that the certificate is compromised}{P1} 

\begin{mytakeaway} Our participants, even though highly technical, demonstrated consistent misunderstandings, with \pct {4} of participants fully understanding all the security questions, indicating that deploying mTLS to everyday users could create serious security risks when problems arise, as users lack the understanding needed for appropriate incident response.
\end{mytakeaway}

%% file: sections/8._limitation.tex
\section{Study Limitations}
\label{sec:limitations}

This study has several limitations that should be considered when interpreting the results. First, written reports risk recall bias \cite{bradburn1987answering, hassan2006recall}. To mitigate this, we triangulated them with ASQ and SUS data \cite{10.5555/2817912.2817913}, providing a more comprehensive view of usability \cite{creswell2007designing}. Secondly, because the study activities counted toward course credit, demand characteristics may have influenced participants’ behavior~\cite{10.1525/collabra.77}. Grades may have made participants more persistent than they would be in everyday use, while the course setting may have encouraged more testing and exploration of edge cases. We therefore cannot tell whether this context made the activities appear easier or harder. To reduce this risk, we emphasized honest feedback and graded written reports only for completion.

Even with course credit precautions, we acknowledge a potential reporting bias in the qualitative data, as participants’ written feedback focused more on negative experiences. However, it is the nature of human to fixate on problems~\cite{baumeister2001bad}. In addition, because the end-of-semester report was collected after the multi device task, we cannot rule out that reflections on routine authentication were influenced by the later migration experience.

Also, our educational setting differed from operational use. Participants interacted with a course server where errors carried fewer consequences than in services such as banking, used raw utilities without prior mTLS-specific training, and did not face the same pressures or receive the organizational support found in professional environments. We also lacked a comparison setup and usage condition with optimized documentation. Consequently, we cannot separate difficulties caused by the instructions from those caused by the tools and platforms. Prior work identifies learnability as a barrier to adopting security mechanisms~\cite{3235866.3235886}, but our results do not establish that the observed problems would occur with the same frequency or severity in other settings.

The participant sample further limits generalizability. Our participants were senior undergraduate and graduate computer science students, not professional developers or security experts. Because of their technical backgrounds, they might have been better and more persistent at troubleshooting than regular users, though we did not test this directly. Since the study was part of a class, students could ask TAs or classmates for help—support that regular users setting up mTLS on their own would not have. Even though TA help was limited and delayed (see \S\ref{procedure:initial_setup}), and we are not sure how much students relied on each other, these resources likely kept them trying after failures that might have caused regular users to give up. As a result, our findings generalize most directly to users with technical training who are not experts. Everyday users may experience greater difficulty, but future work is needed to test this possibility.

Relatedly, our cohort structure raises an additional concern regarding data independence. The two cohorts completed identical tasks, and our analysis found no meaningful differences in their feedback, leading us to analyze them together. However, collaboration or information sharing within or between cohorts may have affected the independence of observations by either making the task easier via shared troubleshooting steps or harder when advice for a different system did not apply. We cannot determine whether such sharing occurred or how it affected the results. Future work should compare genuinely different populations or conditions, such as professional developers and students, or guided and unguided workflows. Finally, although students retained view-only access to their Canvas submissions after the semester, we cannot confirm whether individual students reviewed their past work prior to providing research consent.

%% file: sections/9._discussion.tex
\section{Discussion}
\label{sec:discussion}

\begin{table*}
\centering
\caption{Comparison of credential lifecycle management in prior research and browser based mTLS in this study.}
\label{tab:credential_lifecycle}

\footnotesize
\renewcommand{\arraystretch}{1.12}

\begin{tabularx}{\textwidth}{
    >{\raggedright\arraybackslash}p{0.13\textwidth}
    *{4}{>{\raggedright\arraybackslash}X}
}
\toprule
\textbf{Lifecycle stage}
&
\textbf{PGP and secure email}
&
\textbf{S/MIME}
&
\textbf{IBE style systems}
&
\textbf{Browser based mTLS}
\\

\midrule

Generate credential
&
Easy, few problems \cite{ruoti2015johnny, ruoti2018comparative} 
&
Easy, often handled by IT~\cite{fry2012not} 
&
Easy and automatic \cite{ruoti2018comparative}
&
Difficult, CSR generation is misunderstood and highly error prone
\\

\addlinespace

Register credential
&
Easy, few error cases \cite{bai2016inconvenient} 
&
Easy, usually managed during issuance, although client configuration can be difficult \cite{fry2012not,banday2014s}
&
Easy and automatic \cite{ruoti2018comparative}
&
Extremely difficult because of operating system and browser store fragmentation, poor interfaces, and limited guidance
\\

\addlinespace

Sharing and using credentials
&
Hard, key sharing is fundamentally untenable for most users~\cite{whitten1999johnny,ruoti2019usability}
&

Mixed, depends on IT support and need to configure certificates \cite{garfinkel2005johnny}
&
Easy and automatic, although the protection model may remain unclear \cite{ruoti2013confused,ruoti2018comparative}
&
Easy after enrollment, with no need for user managed sharing. Usage is automatic
\\

\addlinespace

Sync or migration
&
Largely unstudied, with private key backup and transfer left as open problems \cite{ruoti2018comparative}
&
Largely unstudied, although configuration across identities presents related problems \cite{fry2012not,banday2014s}
&
Easy through server mediated key retrieval \cite{ruoti2016private,ruoti2018comparative}
&
Difficult because browsers provide no native identity synchronization or guided export workflow
\\

\bottomrule
\end{tabularx}
\end{table*}


Our findings (\S\ref{sec:results-setup}---\ref{sec:user-understanding}) demonstrate that mTLS has a distinct lifecycle compared to other key-management systems, as summarized in Table~\ref{tab:credential_lifecycle}. We unpack these results in the following subsections and offer recommendations for designing future systems, using mTLS strictly as a motivating example rather than advocating for direct improvements to it.

\subsection{Certificate Creation}
\label{sec:phase_1_discussion}

Our results highlight the gap between generating a key and making a usable certificate. Creating a key pair is simple, yet including it in a valid certificate is difficult (\S\ref{subsection:Key Pair Generation}). We do not offer a fix for this obstacle, but it reveals a wider problem with key management. Users are expected to translate service policies into cryptographic parameters themselves. Thus, key management tools should instead be user-centered and goal-oriented. Rather than asking users to select algorithms, fields, extensions, and formats, a tool should ask what the credential is for, which identity it represents, and whether it will be used on one or multiple devices. The issuer or relying service should provide its policy in a machine-readable form so that the tool can select compatible parameters, validate the credential request, and explain remaining problems in terms of the user's goal. In the studied workflow, this burden included selecting an appropriate key size and certificate fields, validating the CSR, and associating the issued certificate with its private key.

Generally, the system should produce all artifacts needed for registration and explain what was created, where it is stored, which identity and service it represents, and what the user must do next. Our study did not compare candidate creation interfaces, so these are design implications to evaluate rather than established fixes. Nevertheless, evolving TLS requirements illustrate why key-management systems should track technical changes on the user's behalf. For example, current TLS server-authentication standards require service identities to appear in the subject alternative name (SAN) extension and prohibit using the CN for this purpose \cite{saint2023rfc}. A goal-oriented tool could apply such changes without requiring users to monitor and interpret revisions to certificate standards.

\subsection{Key Registration}
\label{sec:phase_2_discussion}

Our findings illustrate a broader registration problem in key-management systems: credential setup is often fragmented across creation tools, storage locations, applications, and service-side validation. Participants had to manage certificate packaging, operating-system storage, browser-specific behavior, and server validation. When interfaces provided no clear confirmation of success, registration errors appeared later as unexplained authentication failures.

These observations support a broader design implication: key-management systems should treat registration as a single, testable operation. Guided by prior work on secondary authentication factor managers \cite{smith2023if}, a registration system should automatically verify that a credential is complete, appropriate for its intended purpose, accessible to the application, and usable with the target service. In mTLS, for example, this would have required confirming that the certificate was associated with its private key, stored where the browser could access it, and capable of authenticating to the intended server. If any check fails, the system should explain the problem and allow the user to retry or safely undo the incomplete registration. This validation principle applies broadly to possession-based credentials, including passkeys and certificate-based credentials.

Broadly, cryptographic credentials should be managed as login identities rather than as unrelated files. In mTLS, for example, this would mean managing client certificates alongside other login credentials instead of requiring users to handle certificate and private-key files separately. Major browsers already include password managers~\cite{silver2014password}, and third-party extensions could offer another way to manage these credentials. Passkeys demonstrate how login credentials can be managed and synchronized through an integrated credential manager \cite{3698900.3699304}. This comparison concerns credential management only. We do not argue that mTLS has an equivalent authentication model, should replace passkeys, or should be prioritized over established passwordless systems.

Automation must also remain transparent. Although automated registration may improve usability, prior studies show that it may leave users uncertain about identity binding, provider control, and key escrow \cite{ruoti2013confused,ruoti2018comparative}. An automated flow should therefore identify the account, service, issuer, storage location, synchronization behavior, and whether the provider retains a copy of the key. Users do not need to understand the underlying cryptography, but they should understand who manages the credential, how recovery works, and what happens if access is lost.

\subsection{Credential Use}
\label{sec:phase_3_discussion}

Routine authentication was easy and straightforward once a client certificate was correctly registered (\S\ref{sec:phase_2}), and participants valued the seamless access provided after setup. This finding supports a broader design principle for key-management systems: cryptographic operations should remain unobtrusive during routine use while the active identity and its scope remain visible. Users should not need to handle raw key material during ordinary authentication, but they should be able to see which identity is active, which service is requesting authentication, and whether authentication succeeded. In our testing, we found that mTLS workflows exposed a broader account management problem. It was difficult to select among matching client certificates and switching identities because these tasks were handled through certificate stores and browser state rather than a clear account interface. Key-management systems should therefore present credentials as named accounts associated with specific services and allow users to switch among them without deleting or reinstalling credentials. Similar account-management problems have affected password managers. Prior work identified limited identity-management support and credential-association failures that made selecting the correct credential difficult \cite{simmons2021systematization,9519389,huaman2024passwords}. The broader implication is that automating routine cryptographic operations should not eliminate users' ability to identify and control the account being used.

\subsection{Migration and Recovery}
\label{sec:phase_4_discussion}

Our findings in \S\ref{sec:phase_3} show that participants did not clearly understand how to manage credentials across multiple devices. 
Because migration and recovery have received limited study, it remains unclear exactly how broadly this problem generalizes. However, we anticipate that similar multi-device usability issues will likely crop up in other systems, and future research should examine these challenges across passkeys and other cryptographic credentials. In our second device task, participants either synchronized an existing credential or issued a separate credential for each device. The available tools did not explain the consequences of these options or provide a supported process for either approach. The broader takeaway is that device management was treated as an improvised file operation rather than as part of the credential lifecycle.

The choice between synchronization and separate device credentials involves important security and usability tradeoffs. Synchronizing one credential reduces enrollment work and can preserve access after device replacement. However, each additional copy creates another location from which the credential could be compromised, and identical copies generally cannot be revoked individually. Separate device credentials support attribution and individual device revocation, but they require reliable enrollment and recovery. Key management systems should explain these tradeoffs when users add a device and provide a suitable default based on the system's threat model and service policy.

Adding a device should be a first class lifecycle operation rather than a manual private key transfer. One possible design pattern is for the new device to generate its own private key and obtain authorization from an enrolled device or another strong recovery channel. A short lived pairing code could connect the two devices without transferring the existing private key. A credential management interface should show all enrolled devices, when each device was added and last used, and which credential it holds. Users should be able to revoke one device without disabling the others. If the same credential is synchronized across devices, the interface should explain that all copies share a single revocation boundary. These capabilities align with the integrated credential management approach discussed under credential use. 

\section{Conclusion}
\label{sec:conclusion}

In a user study of 46 senior and graduate computer science students, we found that mTLS usability varies across the credential lifecycle. Although routine authentication was easy once configured, initial setup, particularly certificate creation and browser registration, proved difficult. Adding a second device introduced unsupported migration workflows and unclear choices between transferring and reissuing credentials. Even after a semester of use, participants rated mTLS poorly and showed limited understanding of its security implications and how to handle key loss, theft, revocation, and migration. These findings extend beyond mTLS. Future key-management systems should guide users in creating valid credentials and integrate registration, device enrollment, recovery, and revocation while keeping routine authentication simple.

%% file: sections/ethics.tex
\section{Ethical Considerations}
\label{sec:ethics}

\subsubsection*{Stakeholders}
The primary stakeholders in this research are the students, whose participation involves potential impacts such as significant time burdens, frustration or stress experienced during failed CSR generation, failed certificate installation, and privacy risks associated with the collection of system logs and free-text reflections. Because the study is situated within an active course context, non-participating students also represent a critical stakeholder group; we ensured that procedures do not inadvertently alter assignments or create unequal burdens that could affect their educational experience. This academic environment further involves the course staff and the university, who bear responsibility for student welfare, grading fairness, and institutional compliance regarding research ethics and the protection of educational records.

Also, browser and OS vendors, PKI tool developers, and enterprise IT teams—are impacted by how these findings are interpreted and whether the resulting recommendations shift industry expectations for client-certificate workflows. Consequently, the wider public and security community may have an interest in this work because it contributes evidence about the usability of mTLS and broader end-user key-management challenges. Finally, potential adversaries and misusers are considered stakeholders in a security context; while the research does not disclose specific software vulnerabilities, the published operational details could, in principle, be repurposed to inform credential theft or social engineering strategies.

\subsubsection*{Research Procedures and Oversight (Respect for Law and Public Interest)}
We embedded the study in a required assignment for an applied cryptography course, designed primarily for pedagogical value by giving students hands-on experience with key management. Research goals did not alter the assignment or grading, and we did not offer compensation. Because the assignment involved configuring certificates on student devices, we consulted with the IRB regarding personal-device use and potential undue burden. University computing facilities reduced, but did not eliminate, the personal-device burden because participants who initially used a university computer were asked to use a personal device for the multi-device task. The setup did not require modifying the trusted-root store or making irreversible operating-system changes. The client certificates were valid only for authentication to the course server. Consequently, leaving residual certificates in a browser posed no security risk to students; even if stolen, the certificates could not be used to alter grades or access unrelated university services. We did not inspect students' devices or collect data directly from them. Our analyses used CA and course-server logs together with participant-written reports. We also provided instructions for removing the course certificate after the assignment.

Also, we obtained IRB approval to use course-generated data for research and, per IRB guidance, sought consent only after final grades were posted to minimize perceived pressure and ensure participation could not affect course standing.

\subsubsection*{Respect for Persons: Autonomy, Consent, and Privacy}

We invited students to participate through a recruitment email linking to an IRB-approved consent form. The form explained the study’s purpose, procedures, participant rights, and our intention to analyze and publish their experiences with TLS client authentication. Students provided consent electronically, confirming that they were at least 18 years old and understood that participation was voluntary and would not affect their grades, relationships with instructors, or university standing. We also provided contact information for the research team and the IRB. 

Students retained view-only access to their Canvas submissions after the semester, allowing them to review, but not modify, their written reports before deciding whether to consent to their use. However, we cannot confirm whether each student reviewed the reports before consenting. Students could decline participation or withdraw consent during the consent collection period. The consent form explained that once analysis of the anonymized data began, individual records could no longer be identified and removed.

We analyzed only data from consenting students, including their written reports from the two key management assignments and relevant CA and course-server logs. We treated all materials as potentially identifying, removed direct identifiers before analysis, and stored the data on access-controlled systems. To prevent re-identification, we limited quotations to non-identifying excerpts and redacted or paraphrased unique device descriptions, unusual mistakes, and self-disclosed personal information.

\subsubsection*{Potential and Mitigation harms during the study.}
Students might experience frustration due to setup difficulty. So, we mitigate this with assignment scaffolding via TA office hours, and clear communication that errors are part of learning; research participation did not affect grades. Also, Setup tasks can consume time beyond typical coursework, so we bounded tasks to course-appropriate scope and monitored for undue burden. Also, participants might adopt poor key management by storing unencrypted private keys, which could create local risk, we warned students not to reuse keys/certs for other services.

\subsubsection*{Justice: Fairness in Recruitment and Burden Distribution}
Students are a convenient population, but course-based research creates risks of inequity and undue
influence. We sought to distribute burdens fairly by ensuring that the course requirement (the assignment)
was pedagogically justified independent of research goals, and by ensuring that opting out of research use
of data did not disadvantage students.

\subsubsection*{Decision to conduct the research} Under a consequentialist analysis, the primary risks (time burden, frustration, and privacy leakage) are low-to-moderate and were mitigated through voluntariness, data minimization, and access controls, while the benefits include improved understanding of a security mechanism with potential real-world impact. Under a deontological analysis, we centered student autonomy through consent and the ability to decline research participation without penalty, and we treated privacy as a right requiring minimization and protection. Considering both lenses, we concluded the study was ethically permissible.

\subsubsection*{Decision to publish.}
Publishing this study carries some risks: readers might apply our findings too broadly, attackers could use our results to exploit systems, or participants could be identified. To prevent this, we clearly state that our findings apply only to our specific educational setting. We also left out technical details that could be misused and paraphrased or redacted potentially identifying excerpts to protect participants' privacy. We believe the benefits of improving key management systems outweigh the small remaining risks.

\section{Generative AI Usage}

We acknowledge the environmental and deskilling impacts of generative AI, including its training on copyrighted texts and the exploitation of workers. We used Grammarly, ChatGPT, and Claude Sonnet~4.6 solely as writing-assistance tools. We wrote the entire paper ourselves in our own words, using these tools only after drafting to check grammar and improve clarity and readability. For ChatGPT and Claude, we provided only isolated, non-consecutive sentences without surrounding context or the broader manuscript. We did not provide complete sections or the entire paper. No tool had access to participant data or was used for data analysis.

%% file: sections/supp.tex
\section{Supplementary Results}
\label{sec:supplementary-results}


\begin{figure}[htbp]
    \centering
    \includegraphics[width=0.99\linewidth]{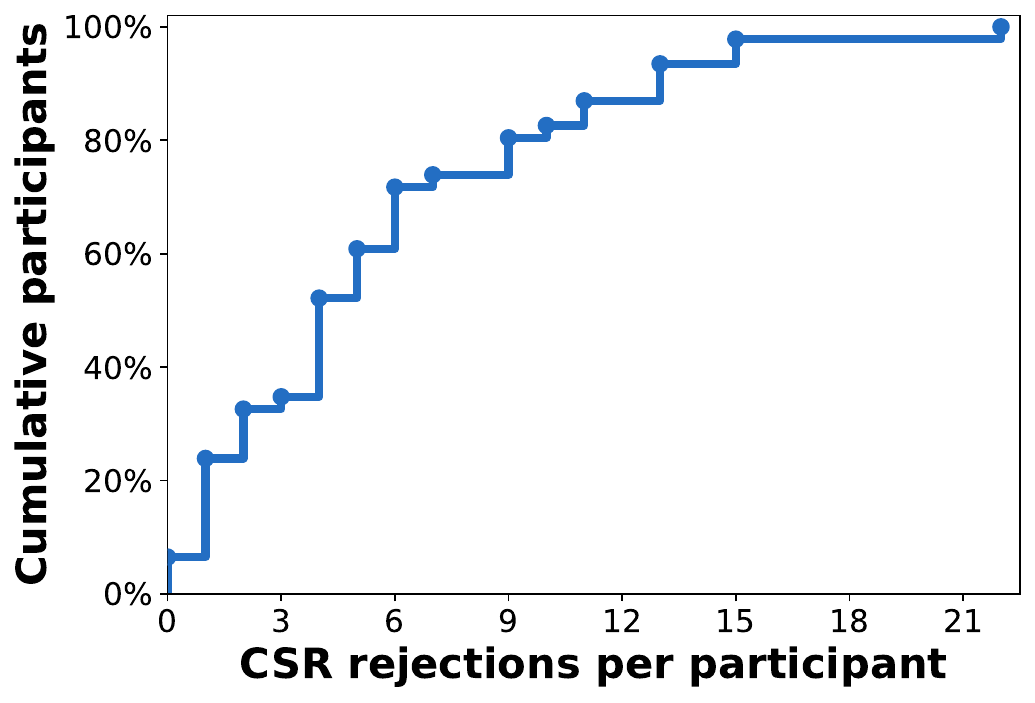}
    \caption{CDF of per-participant counts of rejected CSR}
    \Description{A cumulative distribution of rejected CSR submissions per participant. The median is four rejected submissions, and the maximum is 22.}
    \label{fig:csr-rejection-cdf}
\end{figure}

\begin{table}[htbp]
  \centering
  \footnotesize
  \setlength{\tabcolsep}{3pt}
  \renewcommand{\arraystretch}{1.2}

  \caption{Sensitivity comparison of Cohort 1 with the full aggregated
dataset. The Combined column includes both Cohort 1
and Cohort 2 and shows how key descriptive outcomes changed after
adding Cohort 2. GM denotes geometric mean. All primary analyses use
the combined dataset.}  \label{tab:per-cohort-outcomes}

  \begin{tabular}{@{}p{0.5\columnwidth}rr@{}}
    \toprule
    Outcome & Cohort 1 & Combined \\
    \midrule
    Participants (consented) & 22 & 46 \\
    \midrule
    Initial setup time (self-report; GM, hours) & 2.15 & 2.2 \\
    CSR rejection rate (by request) & 66.4\% (178/268) & 62\% (258/416) \\
    ASQ setup (median, /7) & 2.5 & 3.67 \\
    SUS setup (mean, /100) & 55 & 50 \\
    \midrule
    Second-device setup time (self-report, hours) & 0.39 & 0.48 \\
    ASQ second-device (median, /7) & 5.0 & 5.67 \\
    Migration: generated new credential & 61.9\% & 48\% \\
    Migration: synchronized existing & 38.1\% & 52\% \\
    Used $>$1 certificate during semester & 59\% (13/22) & 80\% (37/46) \\
    End-of-semester SUS (mean, /100) & 45 & 44 \\

    \bottomrule
  \end{tabular}
\end{table}

\begin{figure}
  \centering
  \includegraphics[width=0.99\linewidth]{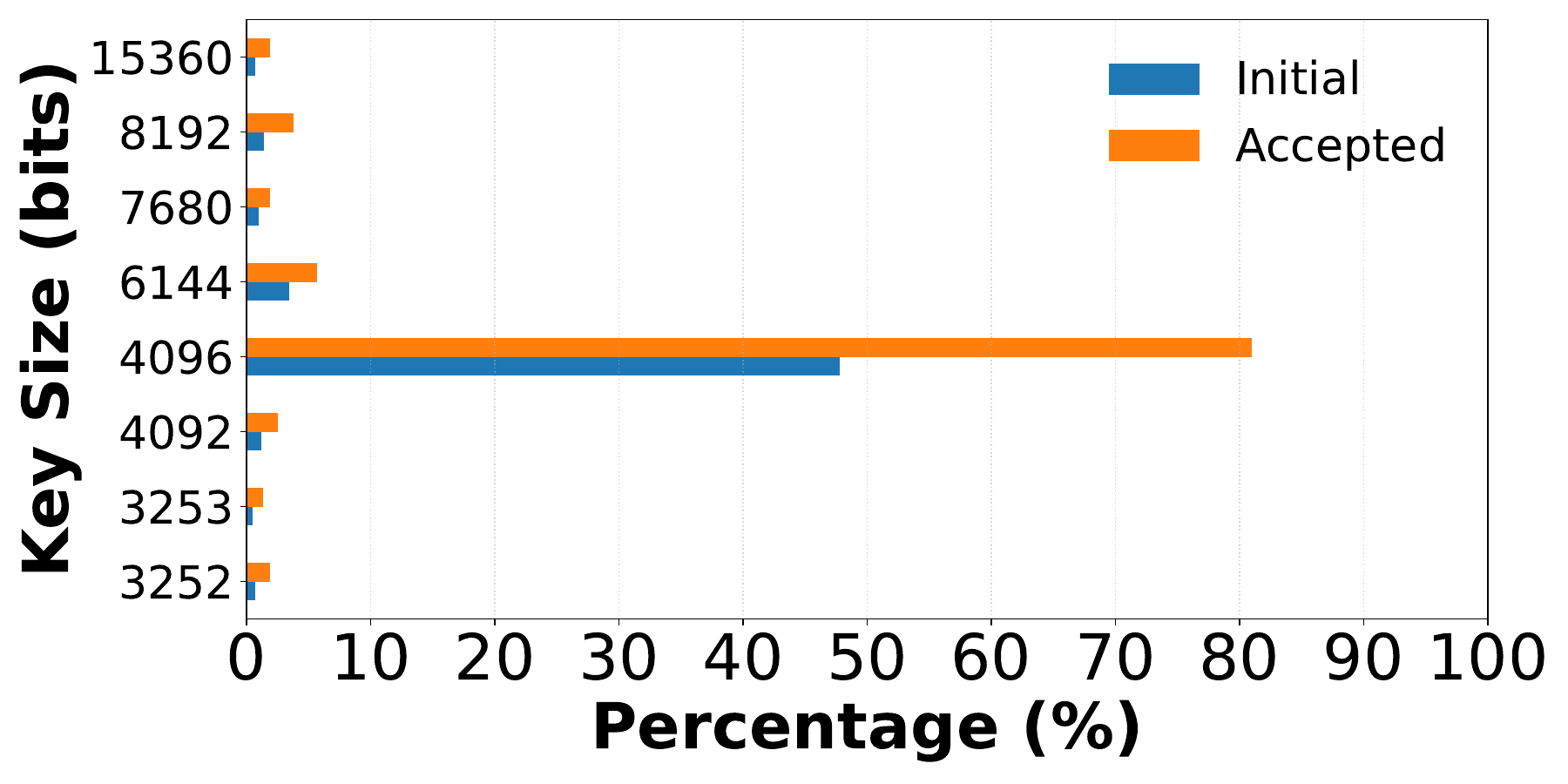}
  \Description{A comparative plot showing the distribution of key lengths
  before and after error feedback, demonstrating a shift toward higher,
  compliant key bit lengths.}
  \caption{Initial versus accepted CSR key-length distributions}
  \label{fig:keysize-shift}
\end{figure}

\begin{figure}
  \centering
  \includegraphics[width=0.99\linewidth]{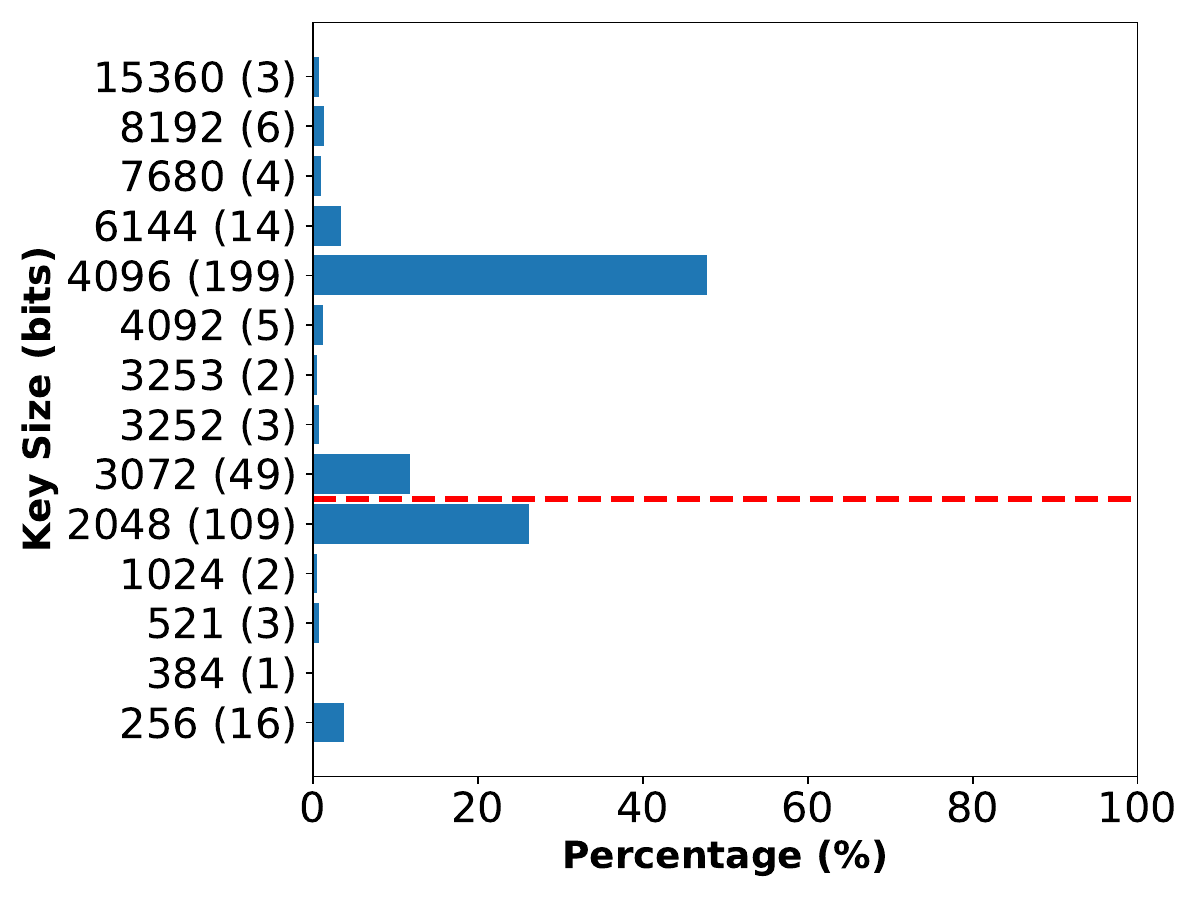}
  \Description{A bar chart displaying the initial key lengths selected by
  users. It highlights that a significant portion of keys were under 3072
  bits and subsequently rejected.}
  \caption{Key-size distribution across all 416 CSR submissions recorded during the study. Counts are shown in parentheses. The dashed line marks the course CA's 3072-bit minimum; submissions below this threshold were rejected.}
  \label{fig:keysize-initial}
\end{figure}

\clearpage

\begin{table*}[!t]
\centering
\footnotesize
\setlength{\tabcolsep}{3pt}
\renewcommand{\arraystretch}{0.92}

\captionof{table}{Consolidated results for 22 exploratory association tests
with Benjamini--Hochberg false discovery rate (BH--FDR) correction.
All 22 association tests were corrected simultaneously at $q=0.05$,
and no association survived correction
($q_{\mathrm{BH}}<.05$). Explanatory and outcome labels indicate
analysis-specific roles only; no variable was experimentally
manipulated, and no causal direction is implied. Effect magnitudes
were classified using unrounded estimates and interpreted as follows:
$|r|$ and $V$, .10 small, .30 medium, and .50 large;
$|d|$, .20 small, .50 medium, and .80 large; and
$\varepsilon^2$, .01 small, .06 medium, and .14
large~\cite{field2024discovering}.}
\label{tab:consolidated-stats}

\begin{tabular}{@{}p{2.8cm}p{2.8cm}lrrrll@{}}
\toprule
\textbf{Explanatory variable}
  & \textbf{Outcome variable}
  & \textbf{Test}
  & \textbf{Statistic}
  & $p$
  & $q_{\mathrm{BH}}$
  & \textbf{Effect size}
  & \textbf{Size} \\
\midrule

\multicolumn{8}{l}{\textit{Phase 1 Setup}} \\
\midrule

CSR errors
  & Initial SUS
  & Spearman
  & $\rho=-0.204$
  & .173 & .517
  & $r=-0.204$
  & small \\

CSR errors
  & Initial ASQ
  & Spearman
  & $\rho=0.198$
  & .188 & .517
  & $r=0.198$
  & small \\

Setup time
  & Initial SUS
  & Spearman
  & $\rho=-0.008$
  & .959 & .988
  & $r=-0.008$
  & negligible \\

Browser failure
  & Initial SUS
  & Welch $t$
  & $t=-2.406$
  & .058 & .317
  & $d=-1.021$
  & large \\

\midrule
\multicolumn{8}{l}{\textit{Phase 3 Multi-Device}} \\
\midrule

Second-device approach (new cert vs.\ sync)
  & Final SUS
  & Welch $t$
  & $t=-0.015$
  & .988 & .988
  & $d=-0.004$
  & negligible \\

Understanding of sync vs.\ new cert
(FC vs.\ NFC)\textsuperscript{a,b}
  & Second-device approach\textsuperscript{a}
  & Pearson $\chi^2$
  & $\chi^2(1)=0.534$
  & .465 & .731
  & $V=0.109$
  & small \\

\midrule
\multicolumn{8}{l}{\textit{Security Understanding and Usability
Outcomes (Kruskal--Wallis, 3 levels)}} \\
\midrule

Q1 mTLS vs.\ password
  & Second-device ASQ
  & KW
  & $H=0.507$
  & .776 & .899
  & $\varepsilon^2=0.000$
  & negligible \\

Q1 mTLS vs.\ password
  & Final SUS
  & KW
  & $H=2.201$
  & .333 & .582
  & $\varepsilon^2=0.005$
  & negligible \\

Q2 sync vs.\ new cert
  & Second-device ASQ
  & KW
  & $H=0.985$
  & .611 & .747
  & $\varepsilon^2=0.000$
  & negligible \\

Q2 sync vs.\ new cert
  & Final SUS
  & KW
  & $H=2.134$
  & .344 & .582
  & $\varepsilon^2=0.003$
  & negligible \\

Q3 lost cert (key intact)
  & Second-device ASQ
  & KW
  & $H=2.582$
  & .275 & .582
  & $\varepsilon^2=0.014$
  & small \\

Q3 lost cert (key intact)
  & Final SUS
  & KW
  & $H=5.051$
  & .080 & .352
  & $\varepsilon^2=0.071$
  & medium \\

Q4 lost private key
  & Second-device ASQ
  & KW
  & $H=10.545$
  & .005 & .113
  & $\varepsilon^2=0.199$
  & large \\

Q4 lost private key
  & Final SUS
  & KW
  & $H=6.697$
  & .035 & .317
  & $\varepsilon^2=0.109$
  & medium \\

Q5 stolen key
  & Second-device ASQ
  & KW
  & $H=2.417$
  & .299 & .582
  & $\varepsilon^2=0.010$
  & negligible \\

Q5 stolen key
  & Final SUS
  & KW
  & $H=2.797$
  & .247 & .582
  & $\varepsilon^2=0.019$
  & small \\

Q6 detection/revocation
  & Second-device ASQ
  & KW
  & $H=0.387$
  & .824 & .907
  & $\varepsilon^2=0.000$
  & negligible \\

Q6 detection/revocation
  & Final SUS
  & KW
  & $H=0.986$
  & .611 & .747
  & $\varepsilon^2=0.000$
  & negligible \\

\midrule
\multicolumn{8}{l}{\textit{Overall Understanding Score
Associations}\textsuperscript{c}} \\
\midrule

\multicolumn{2}{l}{Overall understanding vs.\ CSR errors
  \textsuperscript{a}}
  & Spearman
  & $\rho=-0.101$
  & .504 & .739
  & $r=-0.101$
  & small \\

Overall understanding
  & Final SUS
  & Spearman
  & $\rho=-0.299$
  & .043 & .317
  & $r=-0.299$
  & small \\

Overall understanding
  & Second-device ASQ
  & Spearman
  & $\rho=0.215$
  & .151 & .517
  & $r=0.215$
  & small \\

Overall understanding
  & Initial SUS
  & Spearman
  & $\rho=0.093$
  & .537 & .739
  & $r=0.093$
  & negligible \\

\bottomrule
\end{tabular}

\vspace{1pt}
\begin{minipage}{\textwidth}
\scriptsize
\textsuperscript{a}Symmetric association between two explanatory
attributes; neither variable was assigned an outcome role.

\textsuperscript{b}FC = fully correct; NFC = not fully correct.
For the Pearson chi-square analysis, partially correct and incorrect
responses were combined into the NFC category. The Kruskal--Wallis
analyses retained the original three understanding levels.

\textsuperscript{c}The overall understanding score was the number
of six questions coded as fully correct (range 0--6).

KW = Kruskal--Wallis; $V$ = Cramér's $V$.
For browser failure,
$n_{\mathrm{fail}}=41$ and
$n_{\mathrm{no\text{-}fail}}=5$;
$M_{\mathrm{fail}}=48.08$ and
$M_{\mathrm{no\text{-}fail}}=63.50$.
Negative $\varepsilon^2$ estimates were truncated to 0.000. Assumptions and missing-data procedures are reported in
\S\ref{sub:quantitative-methods}
\end{minipage}

\vspace{0.5em}
\hrule
\vspace{0.5em}

\captionof{table}{Major coded usability challenges and participant-level frequencies.}
\label{tab:challenge-summary}

\centering
\footnotesize 
\setlength{\tabcolsep}{3pt}
\renewcommand{\arraystretch}{0.92} 

\begin{tabularx}{\textwidth}{@{}Xr@{}}
\toprule
\textbf{Major coded challenge} & \textbf{$n/N$ (\%)} \\
\midrule

\multicolumn{2}{@{}l}{\textit{Initial setup}} \\
Browser or OS UI problem                         & 41/46 (89\%) \\
CSR configuration or command syntax              & 37/46 (80\%) \\
Failed or malformed CSR                          & 35/46 (76\%) \\
Key--certificate association / PKCS\#12          & 35/46 (76\%) \\
Repeated trial and recovery                      & 31/46 (67\%) \\
File format, encryption, or password             & 29/46 (63\%) \\
Store discovery and placement                    & 28/46 (61\%) \\
Opaque error or compatibility trap               & 24/46 (52\%) \\
Sparse, conflicting, or outdated information     & 20/46 (43\%) \\

\addlinespace
\multicolumn{2}{@{}l}{\textit{Multi-device management}} \\
New-credential setup problem                     & 11/22 (50\%) \\
Lab, headless, or permission constraint          & 14/46 (30\%) \\
Browser recognition or import problem            & 14/46 (30\%) \\
Transfer or connectivity failure                 & 5/24 (21\%)  \\
Locate or import problem                         & 4/22 (18\%)  \\
Private-key transfer concern                     & 2/46 (4\%)   \\

\bottomrule
\end{tabularx}

\vspace{1pt}
\begin{minipage}{\textwidth}
\footnotesize\textit{Note.}
Initial-setup counts come from the initial written reports, and
multi-device counts come from the second written reports. Counts are
participant-level and non-exclusive; therefore, rows should not be
summed. $N=22$ denotes participants who generated a new credential,
$N=24$ denotes participants who synchronized an existing credential,
and otherwise $N=46$. Percentages are rounded to the nearest whole
number.
\end{minipage}

\end{table*}

%% file: reference.bib
@inproceedings{10.1145/3646547.3688415,
author = {Dong, Hongying and Zhang, Yizhe and Lee, Hyeonmin and Du, Kevin and Tu, Guancheng and Sun, Yixin},
title = {Mutual TLS in Practice: A Deep Dive into Certificate Configurations and Privacy Issues},
year = {2024},
isbn = {9798400705922},
publisher = {Association for Computing Machinery},
address = {New York, NY, USA},
url = {https://doi.org/10.1145/3646547.3688415},
doi = {10.1145/3646547.3688415},
booktitle = {Proceedings of the 2024 ACM Internet Measurement Conference},
pages = {214–229},
numpages = {16},
location = {Madrid, Spain},
series = {IMC '24}
}

@inproceedings{Tiefenau19CertbotUsable,
author = {Tiefenau, Christian and von Zezschwitz, Emanuel and H{\"a}ring, Maximilian and Krombholz, Katharina and Smith, Matthew},
title = {A Usability Evaluation of Let's Encrypt and Certbot: Usable Security Done Right},
year = {2019},
isbn = {9781450367479},
publisher = {Association for Computing Machinery},
address = {New York, NY, USA},
url = {https://doi.org/10.1145/3319535.3363220},
doi = {10.1145/3319535.3363220},
booktitle = {Proceedings of the 2019 ACM SIGSAC Conference on Computer and Communications Security},
pages = {1971–1988},
numpages = {18},
location = {London, United Kingdom},
series = {CCS '19}
}

@article{foppe2018exploiting,
  title={Exploiting tls client authentication for widespread user tracking},
  author={Foppe, Lucas and Martin, Jeremy and Mayberry, Travis and Rye, Erik C and Brown, Lamont},
  journal={Proceedings on Privacy Enhancing Technologies},
  year={2018}
}

@article{parsovs2013practical,
  title={Practical issues with TLS client certificate authentication},
  author={Parsovs, Arnis},
  journal={Cryptology ePrint Archive},
  year={2013}
}

@manual{NGINXSSL,
title     = {NGINX HTTP Server ― SSL Module},
author    = {{NGINX, Inc.}},
year      = {2024},
url       = {https://nginx.org/en/docs/http/ngx_http_ssl_module.html}
}

@article{bangor2008empirical,
  title={An empirical evaluation of the system usability scale},
  author={Bangor, Aaron and Kortum, Philip T and Miller, James T},
  journal={Intl. Journal of Human--Computer Interaction},
  volume={24},
  number={6},
  pages={574--594},
  year={2008},
  publisher={Taylor \& Francis}
}

@article{brooke1996sus,
  title={SUS: A 'Quick and Dirty' Usability Scale},
  author={Brooke, John},
  journal={Usability evaluation in industry},
  volume={189},
  number={3},
  pages={189--194},
  year={1996}
}

@article{10.1145/122672.122692,
author = {Lewis, James R.},
title = {Psychometric evaluation of an after-scenario questionnaire for computer usability studies: the ASQ},
year = {1991},
issue_date = {Jan. 1991},
publisher = {Association for Computing Machinery},
address = {New York, NY, USA},
volume = {23},
number = {1},
issn = {0736-6906},
url = {https://doi.org/10.1145/122672.122692},
doi = {10.1145/122672.122692},
journal = {SIGCHI Bulletin},
month = jan,
pages = {78–81},
numpages = {4}
}

@inproceedings{3698900.3699304,
author = {Lassak, Leona and Pan, Elleen and Ur, Blase and Golla, Maximilian},
title = {Why aren't we using passkeys? obstacles companies face deploying FIDO2 passwordless authentication},
year = {2024},
isbn = {978-1-939133-44-1},
publisher = {USENIX Association},
address = {USA},
booktitle = {Proceedings of the 33rd USENIX Security Symposium},
articleno = {404},
numpages = {18},
location = {Philadelphia, PA, USA},
series = {SEC '24}
}

@INPROCEEDINGS{9511513,
  author={Xia, Wei and Cui, Mingxin and Wang, Wei and Guan, Yangyang and Li, Zhenzhen and Li, Zhen and Xiong, Gang},
  booktitle={2021 28th International Conference on Telecommunications (ICT)}, 
  title={Illuminate the Shadow: A Comprehensive Study of TLS Client Certificate Ecosystem in the Wild}, 
  year={2021},
  volume={},
  number={},
  pages={1-5},
  doi={10.1109/ICT52184.2021.9511513}}

@inproceedings{10.1145/3232755.3232763,
author = {Wachs, Matthias and Scheitle, Quirin and Carle, Georg},
title = {Push Away Your Privacy: Precise User Tracking Based on TLS Client Certificate Authentication},
year = {2018},
isbn = {9781450355858},
publisher = {Association for Computing Machinery},
address = {New York, NY, USA},
url = {https://doi.org/10.1145/3232755.3232763},
doi = {10.1145/3232755.3232763},
booktitle = {Proceedings of the 2018 Applied Networking Research Workshop},
pages = {3},
numpages = {1},
location = {Montreal, QC, Canada},
series = {ANRW '18}
}

@inproceedings{reynolds2018tale,
  title={A tale of two studies: The best and worst of yubikey usability},
  author={Reynolds, Joshua and Smith, Trevor and Reese, Ken and Dickinson, Luke and Ruoti, Scott and Seamons, Kent},
  booktitle={2018 IEEE symposium on security and privacy (SP)},
  pages={872--888},
  year={2018},
  organization={IEEE}
}

@inproceedings{3563572.3563576,
author = {Owens, Kentrell and Anise, Olabode and Krauss, Amanda and Ur, Blase},
title = {User perceptions of the usability and security of smartphones as FIDO2 roaming authenticators},
year = {2021},
isbn = {978-1-939133-25-0},
publisher = {USENIX Association},
address = {USA},
booktitle = {Proceedings of the Seventeenth Symposium on Usable Privacy and Security},
articleno = {4},
numpages = {20},
series = {SOUPS'21}
}

@misc{cryptography,
title        = {cryptography},
author       = {The {cryptography} developers},
year         = {2025},
howpublished = {\url{https://cryptography.io/}},
}

@article{hassan2006recall,
  title={Recall bias can be a threat to retrospective and prospective research designs},
  author={Hassan, Eman Mohammed Hassan},
  journal={The internet journal of epidemiology},
  year={2006},
    volume={3},
  number={2},
  url={https://ispub.com/ije/3/2/13060}
}

@article{bradburn1987answering,
  title={Answering autobiographical questions: The impact of memory and inference on surveys},
  author={Bradburn, Norman M and Rips, Lance J and Shevell, Steven K},
  journal={Science},
  volume={236},
  number={4798},
  pages={157--161},
  year={1987},
  publisher={American Association for the Advancement of Science}
}

@INPROCEEDINGS{9152694,
  author={Ghorbani Lyastani, Sanam and Schilling, Michael and Neumayr, Michaela and Backes, Michael and Bugiel, Sven},
  booktitle={2020 IEEE Symposium on Security and Privacy (SP)}, 
  title={Is FIDO2 the Kingslayer of User Authentication? A Comparative Usability Study of FIDO2 Passwordless Authentication}, 
  year={2020},
  volume={},
  number={},
  pages={268-285},
  doi={10.1109/SP40000.2020.00047}}

@book{creswell2007designing,
author    = {John W. Creswell and Vicki L. Plano Clark},
title     = {Designing and Conducting Mixed Methods Research},
edition   = {3rd},
year      = {2017},
publisher = {SAGE Publications},
address   = {Thousand Oaks, CA}
}

@manual{openssl_genrsa,
title = {OpenSSL genrsa Manual Page},
organization = {OpenSSL Software Foundation},
note = {Version 1.1.1},
url = {https://www.openssl.org/docs/man1.1.1/man1/genrsa.html},
year = {2024}
}

@online{gitguardian,
author       = {{GitGuardian}},
title        = {GitGuardian: Secrets Security Platform},
year         = {2025},
url          = {https://www.gitguardian.com},
note         = {Accessed: 2025-10-30}
}

@INPROCEEDINGS{6234436,
  author={Bonneau, Joseph and Herley, Cormac and Oorschot, Paul C. van and Stajano, Frank},
  booktitle={2012 IEEE Symposium on Security and Privacy}, 
  title={The Quest to Replace Passwords: A Framework for Comparative Evaluation of Web Authentication Schemes}, 
  year={2012},
  volume={},
  number={},
  pages={553-567},
  doi={10.1109/SP.2012.44}}

@inproceedings{farke2020you,
author = {Farke, Florian M. and Lorenz, Lennart and Schnitzler, Theodor and Markert, Philipp and D{\"u}rmuth, Markus},
title = {"You still use the password after all" — exploring FIDO2 security keys in a small company},
year = {2020},
isbn = {978-1-939133-16-8},
publisher = {USENIX Association},
address = {USA},
booktitle = {Proceedings of the Sixteenth Symposium on Usable Privacy and Security},
articleno = {2},
numpages = {17},
pages={19--35},
series = {SOUPS'20}
}

@inproceedings{3698900.3699303,
author = {Fischer, Konstantin and Trummov{\'a}, Ivana and Gajland, Phillip and Acar, Yasemin and Fahl, Sascha and Sasse, Angela},
title = {The challenges of bringing cryptography from research papers to products: results from an interview study with experts},
year = {2024},
isbn = {978-1-939133-44-1},
publisher = {USENIX Association},
address = {USA},
booktitle = {Proceedings of the 33rd USENIX Security Symposium},
articleno = {403},
numpages = {18},
location = {Philadelphia, PA, USA},
series = {SEC '24}
}

@book{field2024discovering,
  title={Discovering Statistics Using IBM SPSS Statistics},
  author={Field, Andy},
  year={2024},
  publisher={SAGE Publications Ltd},
  edition={6th}
}

@article{10.1525/collabra.77,
    author = {Bowen, Holly J. and Kensinger, Elizabeth A.},
    title = {Cash or Credit? Compensation in Psychology Studies: Motivation Matters},
    journal = {Collabra: Psychology},
    volume = {3},
    number = {1},
    pages = {12},
    year = {2017},
    month = {05},
    issn = {2474-7394},
    doi = {10.1525/collabra.77},
    url = {https://doi.org/10.1525/collabra.77},
    eprint = {https://online.ucpress.edu/collabra/article-pdf/3/1/12/467166/77-944-2-pb.pdf},
}

@article{barker2020nist,
title={NIST SP 800-57 Part 1 Rev. 5: Recommendation for Key Management: Part 1--General},
author={Barker, Elaine},
journal={NIST Standard},
year={2020},
url= {https://csrc.nist.gov/publications/detail/sp/800-57-part-1/rev-5/final},

}

@techreport{barker2018recommendation,
  title={Recommendation for key management, part 2: best practices for key management organization},
  author={Barker, Elaine and Barker, William},
  year={2018},
  institution={National Institute of Standards and Technology}
}

@techreport{barker2018transitioning,
  title={Transitioning the use of cryptographic algorithms and key lengths},
  author={Barker, Elaine and Roginsky, Allen},
  year={2018},
  institution={National Institute of Standards and Technology}
}

@article{10.5555/2817912.2817913,
author = {Brooke, John},
title = {SUS: a retrospective},
year = {2013},
issue_date = {February 2013},
publisher = {Usability Professionals' Association},
address = {Bloomingdale, IL},
volume = {8},
number = {2},
journal = {J. Usability Studies},
month = feb,
pages = {29–40},
numpages = {12}
}

@inproceedings{3241189.3241293,
author = {Krombholz, Katharina and Mayer, Wilfried and Schmiedecker, Martin and Weippl, Edgar},
title = {"I have no idea what i'm doing": on the usability of deploying HTTPS},
year = {2017},
isbn = {9781931971409},
publisher = {USENIX Association},
address = {USA},
booktitle = {Proceedings of the 26th USENIX Security Symposium},
pages = {1339–1356},
numpages = {18},
location = {Vancouver, BC, Canada},
series = {SEC'17}
}

@article{10.1109/MSP.2004.71,
author = {Balfanz, Dirk and Durfee, Glenn and Grinter, Rebecca E. and Smetters, D. K.},
title = {In Search of Usable Security: Five Lessons from the Field},
year = {2004},
issue_date = {September 2004},
publisher = {IEEE Educational Activities Department},
address = {USA},
volume = {2},
number = {5},
issn = {1540-7993},
url = {https://doi.org/10.1109/MSP.2004.71},
doi = {10.1109/MSP.2004.71},
journal = {IEEE Security and Privacy},
month = sep,
pages = {19–24},
numpages = {6}
}

@article{10.1145/3419472,
author = {Ukrop, Martin and Kraus, Lydia and Matyas, Vashek},
title = {Will You Trust This TLS Certificate? Perceptions of People Working in IT (Extended Version)},
year = {2020},
issue_date = {December 2020},
publisher = {Association for Computing Machinery},
address = {New York, NY, USA},
volume = {1},
number = {4},
url = {https://doi.org/10.1145/3419472},
doi = {10.1145/3419472},
journal = {Digital Threats},
month = dec,
articleno = {25},
numpages = {29}
}

@INPROCEEDINGS{8835228,
  author={Krombholz, Katharina and Busse, Karoline and Pfeffer, Katharina and Smith, Matthew and von Zezschwitz, Emanuel},
  booktitle={2019 IEEE Symposium on Security and Privacy (SP)}, 
  title={"If HTTPS Were Secure, I Wouldn't Need 2FA" - End User and Administrator Mental Models of HTTPS}, 
  year={2019},
  volume={},
  number={},
  pages={246-263},
  doi={10.1109/SP.2019.00060}}

@inproceedings{3291228.3291260,
author = {Wu, Justin and Zappala, Daniel},
title = {When is a tree really a truck? exploring mental models of encryption},
year = {2018},
isbn = {9781931971454},
publisher = {USENIX Association},
address = {USA},
booktitle = {Proceedings of the Fourteenth Symposium on Usable Privacy and Security},
pages = {395–409},
numpages = {15},
location = {Baltimore, MD, USA},
series = {SOUPS '18}
}

@inproceedings{3235866.3235886,
author = {Khan, Hassan and Hengartner, Urs and Vogel, Daniel},
title = {Usability and security perceptions of implicit authentication: convenient, secure, sometimes annoying},
year = {2015},
isbn = {9781931971249},
publisher = {USENIX Association},
address = {USA},
booktitle = {Proceedings of the Eleventh Symposium on Usable Privacy and Security},
pages = {225–239},
numpages = {15},
location = {Ottawa, Canada},
series = {SOUPS '15}
}

@inproceedings {219400,
author = {Haney, Julie M. and Theofanos, Mary F. and Acar, Yasemin and Prettyman, Sandra Spickard},
title = {"We make it a big deal in the company": security mindsets in organizations that develop cryptographic products},
year = {2018},
isbn = {9781931971454},
publisher = {USENIX Association},
address = {USA},
booktitle = {Proceedings of the Fourteenth Symposium on Usable Privacy and Security},
pages = {357–373},
numpages = {17},
location = {Baltimore, MD, USA},
series = {SOUPS '18}
}

@article{ellison2000ten,
  title={Ten risks of PKI: What you're not being told about public key infrastructure},
  author={Ellison, Carl and Schneier, Bruce},
  journal={Comput Secur J},
  volume={16},
  number={1},
  pages={1--7},
  year={2000}
}

@article{smith1994misconceptions,
  title={Misconceptions reconceived: A constructivist analysis of knowledge in transition},
  author={Smith III, John P and DiSessa, Andrea A and Roschelle, Jeremy},
  journal={The journal of the learning sciences},
  volume={3},
  number={2},
  pages={115--163},
  year={1994},
  publisher={Taylor \& Francis}
}

@inproceedings{kaczmarczyk2010identifying,
author = {Kaczmarczyk, Lisa C. and Petrick, Elizabeth R. and East, J. Philip and Herman, Geoffrey L.},
title = {Identifying student misconceptions of programming},
year = {2010},
isbn = {9781450300063},
publisher = {Association for Computing Machinery},
address = {New York, NY, USA},
url = {https://doi.org/10.1145/1734263.1734299},
doi = {10.1145/1734263.1734299},
booktitle = {Proceedings of the 41st ACM Technical Symposium on Computer Science Education},
pages = {107–111},
numpages = {5},
location = {Milwaukee, Wisconsin, USA},
series = {SIGCSE '10}
}

@techreport{rescorla2018transport,
  title={The transport layer security (TLS) protocol version 1.3},
  author={Rescorla, Eric},
  year={2018},
  doi={10.17487/RFC8446}
}

@techreport{nystrom2000pkcs,
  title={RFC2986: PKCS\# 10: Certification request syntax specification version 1.7},
  author={Nystrom, Magnus and Kaliski, Burt},
  year={2000},
  doi         = {10.17487/RFC2986}
}

@misc{cooper2008internet,
author={Cooper, David and Santesson, Stefan and Farrell, Stephen and Boeyen, Sharon and Housley, Rusell and Polk, William},
title = {RFC 5280: Internet X.509 Public Key Infrastructure Certificate and Certificate Revocation List (CRL) Profile},
year = {2008},
publisher = {RFC Editor},
address = {USA}
}

@misc{10.17487/RFC7292,
author = {Nystrom, M. and Parkinson, S. and Rusch, A. and Scott, M.},
editor = {K. Moriarty},
title = {RFC 7292: PKCS \#12: Personal Information Exchange Syntax v1.1},
year = {2014},
publisher = {RFC Editor},
address = {USA}
}

@inproceedings{10.1145/2974927.2974938,
author = {Uda, Satoshi and Shikida, Mikifumi},
title = {Challenges of Deploying PKI based Client Digital Certification},
year = {2016},
isbn = {9781450340953},
publisher = {Association for Computing Machinery},
address = {New York, NY, USA},
url = {https://doi.org/10.1145/2974927.2974938},
doi = {10.1145/2974927.2974938},
booktitle = {Proceedings of the 2016 ACM SIGUCCS Annual Conference},
pages = {55–60},
numpages = {6},
location = {Denver, Colorado, USA},
series = {SIGUCCS '16}
}

@inproceedings{conners2022let,
  title={Let’s authenticate: Automated certificates for user authentication},
  author={Conners, James and Devenport, Corey and Derbidge, Stephen and Farnsworth, Natalie and Gates, Kyler and Lambert, Stephen and McClain, Christopher and Nichols, Parker and Zappala, Daniel},
  booktitle={Network and Distributed System Security Symposium (NDSS)},
  year={2022}
}

@misc{franken_2026_18246555,
  author       = {Franken, Gertjan and
                  Vanderlinden, Vik and
                  Tiwari, Vinod and
                  Duggal, Anirudh and
                  Kraus, Martina and
                  Franken, Gertjan and
                  Clark, Brian and
                  Rautenstrauch, Jannis and
                  Daan, Vansteenhuyse and
                  Tiwari, Abhishek},
  title        = {Security},
  month        = jan,
  year         = 2026,
  publisher    = {HTTP Archive},
  doi          = {10.5281/zenodo.18246555},
  url          = {https://doi.org/10.5281/zenodo.18246555},
}

@inproceedings {whitten1999johnny,
	author = {Alma Whitten and J. D. Tygar},
	title = {Why Johnny Can{\textquoteright}t Encrypt: A Usability Evaluation of {PGP} 5.0},
	booktitle = {8th USENIX Security Symposium (USENIX Security 99)},
	year = {1999},
	address = {Washington, D.C.},
	url = {https://www.usenix.org/conference/8th-usenix-security-symposium/why-johnny-cant-encrypt-usability-evaluation-pgp-50},
	publisher = {USENIX Association},
	month = aug
}

@article{ruoti2015johnny,
  title={Why Johnny still, still can't encrypt: Evaluating the usability of a modern PGP client},
  author={Ruoti, Scott and Andersen, Jeff and Zappala, Daniel and Seamons, Kent},
  journal={arXiv preprint arXiv:1510.08555},
  year={2015}
}

@inproceedings{garfinkel2005johnny,
author = {Garfinkel, Simson L. and Miller, Robert C.},
title = {Johnny 2: a user test of key continuity management with S/MIME and Outlook Express},
year = {2005},
isbn = {1595931783},
publisher = {Association for Computing Machinery},
address = {New York, NY, USA},
url = {https://doi.org/10.1145/1073001.1073003},
doi = {10.1145/1073001.1073003},
booktitle = {Proceedings of the 2005 Symposium on Usable Privacy and Security},
pages = {13–24},
numpages = {12},
location = {Pittsburgh, Pennsylvania, USA},
series = {SOUPS '05}
}

@inproceedings{stransky202227,
  author={Stransky, Christian and Wiese, Oliver and Roth, Volker and Acar, Yasemin and Fahl, Sascha},
  booktitle={2022 IEEE Symposium on Security and Privacy (SP)}, 
  title={27 Years and 81 Million Opportunities Later: Investigating the Use of Email Encryption for an Entire University}, 
  year={2022},
  volume={},
  number={},
  pages={860-875},
  doi={10.1109/SP46214.2022.9833755}}

@inproceedings{ruoti2016private,
author = {Ruoti, Scott and Andersen, Jeff and Hendershot, Travis and Zappala, Daniel and Seamons, Kent},
title = {Private Webmail 2.0: Simple and Easy-to-Use Secure Email},
year = {2016},
isbn = {9781450341899},
publisher = {Association for Computing Machinery},
address = {New York, NY, USA},
url = {https://doi.org/10.1145/2984511.2984580},
doi = {10.1145/2984511.2984580},
booktitle = {Proceedings of the 29th Annual Symposium on User Interface Software and Technology},
pages = {461–472},
numpages = {12},
location = {Tokyo, Japan},
series = {UIST '16}
}

@inproceedings{ruoti2018comparative,
author = {Ruoti, Scott and Andersen, Jeff and Monson, Tyler and Zappala, Daniel and Seamons, Kent},
title = {A comparative usability study of key management in secure email},
year = {2018},
isbn = {9781931971454},
publisher = {USENIX Association},
address = {USA},
booktitle = {Proceedings of the Fourteenth Symposium on Usable Privacy and Security},
pages = {375–394},
numpages = {20},
location = {Baltimore, MD, USA},
series = {SOUPS '18}
}

@inproceedings{bai2016inconvenient,
author = {Bai, Wei and Kim, Doowon and Namara, Moses and Qian, Yichen and Kelley, Patrick Gage and Mazurek, Michelle L.},
title = {An inconvenient trust: user attitudes toward security and usability tradeoffs for key-directory encryption systems},
year = {2016},
isbn = {9781931971317},
publisher = {USENIX Association},
address = {USA},
booktitle = {Proceedings of the Twelfth Symposium on Usable Privacy and Security},
pages = {113–130},
numpages = {18},
location = {Denver, CO, USA},
series = {SOUPS '16}
}

@article{ruoti2019usability,
author = {Ruoti, Scott and Andersen, Jeff and Dickinson, Luke and Heidbrink, Scott and Monson, Tyler and O'neill, Mark and Reese, Ken and Spendlove, Brad and Vaziripour, Elham and Wu, Justin and Zappala, Daniel and Seamons, Kent},
title = {A Usability Study of Four Secure Email Tools Using Paired Participants},
year = {2019},
issue_date = {May 2019},
publisher = {Association for Computing Machinery},
address = {New York, NY, USA},
volume = {22},
number = {2},
issn = {2471-2566},
url = {https://doi.org/10.1145/3313761},
doi = {10.1145/3313761},
journal = {ACM Transactions on Privacy and Security (TOPS)},
month = apr,
articleno = {13},
numpages = {33}
}

@inproceedings{ruoti2016we,
author = {Ruoti, Scott and Andersen, Jeff and Heidbrink, Scott and O'Neill, Mark and Vaziripour, Elham and Wu, Justin and Zappala, Daniel and Seamons, Kent},
title = {"We're on the Same Page": A Usability Study of Secure Email Using Pairs of Novice Users},
year = {2016},
isbn = {9781450333627},
publisher = {Association for Computing Machinery},
address = {New York, NY, USA},
url = {https://doi.org/10.1145/2858036.2858400},
doi = {10.1145/2858036.2858400},
booktitle = {Proceedings of the 2016 CHI Conference on Human Factors in Computing Systems},
pages = {4298–4308},
numpages = {11},
location = {San Jose, California, USA},
series = {CHI '16}
}

@inproceedings{ruoti2013confused,
  author = {Ruoti, Scott and Kim, Nathan and Burgon, Ben and van der Horst, Timothy and Seamons, Kent},
  title = {Confused Johnny: when automatic encryption leads to confusion and mistakes},
  year = {2013},
  isbn = {9781450323192},
  publisher = {Association for Computing Machinery},
  address = {New York, NY, USA},
  url = {https://doi.org/10.1145/2501604.2501609},
  doi = {10.1145/2501604.2501609},
  booktitle = {Proceedings of the Ninth Symposium on Usable Privacy and Security},
  articleno = {5},
  numpages = {12},
  location = {Newcastle, United Kingdom},
  pages={1--12},
  series = {SOUPS '13}
}

@article{borradaile2021motivated,
  author={Borradaile, Glencora and Kretschmer, Kelsy and Gretes, Michele and LeClerc, Alexandria},
  title   = {The Motivated Can Encrypt (Even with {PGP})},
  journal = {Proceedings on Privacy Enhancing Technologies},
  volume  = {2021},
  number  = {3},
  pages   = {49--69},
  year    = {2021},
  doi     = {10.2478/popets-2021-0037},
  url     = {https://doi.org/10.2478/popets-2021-0037}
}

@article{ruoti2019johnny,
  author={Ruoti, Scott and Seamons, Kent},
  journal={IEEE Security \& Privacy}, 
  title={Johnny's Journey Toward Usable Secure Email}, 
  year={2019},
  volume={17},
  number={6},
  pages={72-76},
  doi={10.1109/MSEC.2019.2933683}}

@inproceedings{holtervennhoff2024mixed,
author = {Sandra H{\"o}ltervennhoff and Noah W{\"o}hler and Arne M{\"o}hle and Marten Oltrogge and Yasemin Acar and Oliver Wiese and Sascha Fahl},
title = {A {Mixed-Methods} Study on User Experiences and Challenges of Recovery Codes for an {End-to-End} Encrypted Service},
booktitle = {33rd USENIX Security Symposium (USENIX Security 24)},
year = {2024},
isbn = {978-1-939133-44-1},
address = {Philadelphia, PA},
pages = {7267--7284},
publisher = {USENIX Association},
month = aug
}

@inproceedings{daffalla2025framework,
author = {Daffalla, Alaa and Bhattacharya, Arkaprabha and Wilder, Jacob and Chatterjee, Rahul and Dell, Nicola and Bellini, Rosanna and Ristenpart, Thomas},
title = {A framework for abusability analysis: the case of passkeys in interpersonal threat models},
year = {2025},
isbn = {978-1-939133-52-6},
booktitle = {Proceedings of the 34th USENIX Security Symposium},
articleno = {401},
numpages = {20},
location = {Seattle, WA, USA},
  pages={7819--7838},
series = {SEC '25}
}

@inproceedings{fry2012not,
  title={Not sealed but delivered: The (un) usability of s/mime today},
  author={Fry, Ann and Chiasson, Sonia and Somayaji, Anil},
  booktitle={Annual Symposium on Information Assurance and Secure Knowledge Management (ASIA’12), Albany, NY},
  year={2012}
}

@inproceedings{banday2014s,
  title={S/MIME with multiple e-mail address certificates: A usability study},
  author={Banday, M Tariq and Sheikh, Shafiya Afzal},
  booktitle={2014 International Conference on Contemporary Computing and Informatics (IC3I)},
  pages={707--712},
  year={2014},
  organization={IEEE}
}

@misc{saint2023rfc,
author = {Saint-Andre, P. and Salz, R.},
title = {RFC 9525: Service Identity in TLS},
year = {2023},
publisher = {RFC Editor},
address = {USA},
url={https://hjp.at/doc/rfc/rfc9525.pdf}
}

@inproceedings{smith2023if,
author = {Smith, Garrett and Yadav, Tarun and Dutson, Jonathan and Ruoti, Scott and Seamons, Kent},
title = {"If I could do this, I feel anyone could:": the design and evaluation of a secondary authentication factor manager},
year = {2023},
isbn = {978-1-939133-37-3},
publisher = {USENIX Association},
address = {USA},
booktitle={32nd USENIX Security Symposium (USENIX Security 23)},
articleno = {29},
numpages = {18},
location = {Anaheim, CA, USA},
pages={499--515},
series = {SEC '23}
}

@inproceedings{huaman2024passwords,
  title={Passwords To-Go: Investigating Multifaceted Challenges for Password Managers in the Android Ecosystem},
  author={Huaman, Nicolas and Oltrogge, Marten and Klivan, Sabrina and Evers, Yannick and Fahl, Sascha},
  booktitle={2024 Annual Computer Security Applications Conference (ACSAC)},
  pages={304--320},
  year={2024},
  organization={IEEE}
}

@inproceedings{simmons2021systematization,
  title={Systematization of password manager use cases and design paradigms},
  author={Simmons, James and Diallo, Oumar and Oesch, Sean and Ruoti, Scott},
  booktitle={Proceedings of the 37th Annual Computer Security Applications Conference},
  pages={528--540},
  year={2021}
}

@INPROCEEDINGS{9519389,
  author={Huaman, Nicolas and Klivan, Sabrina and Oltrogge, Marten and Acar, Yasemin and Fahl, Sascha},
  booktitle={2021 IEEE Symposium on Security and Privacy (SP)}, 
  title={They Would do Better if They Worked Together: The Case of Interaction Problems Between Password Managers and Websites}, 
  year={2021},
  volume={},
  number={},
  pages={1367-1381},
  doi={10.1109/SP40001.2021.00094}}

@article{holton2007coding,
  title={The coding process and its challenges},
  author={Holton, Judith A},
  journal={The Sage handbook of grounded theory},
  volume={3},
  pages={265--289},
  year={2007}
}

@article{glaser1965constant,
  title={The constant comparative method of qualitative analysis},
  author={Glaser, Barney G},
  journal={Social problems},
  volume={12},
  number={4},
  pages={436--445},
  year={1965},
  publisher={Oxford University Press Oxford, UK}
}

@inproceedings{silver2014password,
author = {Silver, David and Jana, Suman and Boneh, Dan and Chen, Eric and Jackson, Collin},
title = {Password managers: attacks and defenses},
year = {2014},
isbn = {9781931971157},
publisher = {USENIX Association},
address = {USA},
booktitle = {Proceedings of the 23rd USENIX Security Symposium},
pages = {449–464},
numpages = {16},
location = {San Diego, CA},
series = {SEC'14}
}

@article{benjamini1995controlling,
  title={Controlling the false discovery rate: a practical and powerful approach to multiple testing},
  author={Benjamini, Yoav and Hochberg, Yosef},
  journal={Journal of the Royal statistical society: series B (Methodological)},
  volume={57},
  number={1},
  pages={289--300},
  year={1995},
  publisher={Wiley Online Library}
}

@article{holm1979simple,
  title={A simple sequentially rejective multiple test procedure},
  author={Holm, Sture},
  journal={Scandinavian journal of statistics},
  pages={65--70},
  year={1979},
  publisher={JSTOR}
}

@article{welch1947generalization,
  title={The generalisation of student's problems when several different population variances are involved},
  author={Welch, Bernard L},
  journal={Biometrika},
  volume={34},
  number={1-2},
  pages={28--35},
  year={1947},
  publisher={Oxford University Press}
}

@article{kruskal1952use,
  title={Use of ranks in one-criterion variance analysis},
  author={Kruskal, William H and Wallis, W Allen},
  journal={Journal of the American statistical Association},
  volume={47},
  number={260},
  pages={583--621},
  year={1952},
  publisher={Taylor \& Francis}
}

@article{mann1947test,
  title={On a test of whether one of two random variables is stochastically larger than the other},
  author={Mann, Henry B and Whitney, Donald R},
  journal={The annals of mathematical statistics},
  pages={50--60},
  year={1947},
  publisher={JSTOR}
}

@article{friedman1937use,
  title={The use of ranks to avoid the assumption of normality implicit in the analysis of variance},
  author={Friedman, Milton},
  journal={Journal of the american statistical association},
  volume={32},
  number={200},
  pages={675--701},
  year={1937},
  publisher={Taylor \& Francis}
}

@article{baumeister2001bad,
  title={Bad is stronger than good},
  author={Baumeister, Roy F and Bratslavsky, Ellen and Finkenauer, Catrin and Vohs, Kathleen D},
  journal={Review of general psychology},
  volume={5},
  number={4},
  pages={323--370},
  year={2001},
  publisher={SAGE Publications Sage CA: Los Angeles, CA}
}

@inproceedings{bhardwaj2026state,
author = {Bhardwaj, Prince and Sastry, Nishanth},
  title={State of Passkey Authentication in the Wild: A Census of the Top 100K sites},
year = {2026},
isbn = {978-3-032-18267-8},
publisher = {Springer-Verlag},
address = {Berlin, Heidelberg},
url = {https://doi.org/10.1007/978-3-032-18268-5_15},
doi = {10.1007/978-3-032-18268-5_15},
booktitle={International Conference on Passive and Active Network Measurement},
pages = {319–347},
numpages = {29}
}

@techreport{temoshok2024digital,
  title={Digital identity guidelines: Authentication and authenticator management},
  author={Temoshok, David and Fenton, James and Choong, Yee-Yin and Lefkovitz, Naomi and Regenscheid, Andrew and Richer, Justin},
  year={2024},
  institution={National Institute of Standards and Technology},
  doi={https://doi.org/10.6028/NIST.SP.800-63b-4}
}

@misc{aumasson_murphy,
  author       = {Jean-Philippe Aumasson},
  title        = {Murphy's laws of cryptography},
  howpublished = {\url{https://www.aumasson.jp/murphy.html}},
}

@inproceedings{zimmermann12018quest,
    author       = {Zimmermann, Verena and Gerber, Nina and Kleboth, Marius and von Preuschen, Alexandra and Schmidt, Konstantin and Mayer, Peter},
    year         = {2018},
    title        = {The Quest to Replace Passwords Revisited - Rating Authentication Schemes},
    pages        = {38-48},
    booktitle    = {Proceedings of the 12th International Symposium on Human Aspects of Information Security and Assurance},
    isbn         = {978-0-244-40254-9},
    language     = {english}
}

@article{marky2022nah,
  title={“Nah, it’s just annoying!” A deep dive into user perceptions of two-factor authentication},
  author={Marky, Karola and Ragozin, Kirill and Chernyshov, George and Matviienko, Andrii and Schmitz, Martin and M{\"u}hlh{\"a}user, Max and Eghtebas, Chloe and Kunze, Kai},
  journal={ACM transactions on computer-human interaction},
  volume={29},
  number={5},
  pages={1--32},
  year={2022},
  publisher={ACM New York, NY}
}

@inproceedings{lassak2021s,
author = {Leona Lassak and Annika Hildebrandt and Maximilian Golla and Blase Ur},
title = {"It{\textquoteright}s Stored, Hopefully, on an Encrypted Server{\textquoteright}{\textquoteright}: Mitigating Users{\textquoteright} Misconceptions About {FIDO2} Biometric {WebAuthn}},
booktitle = {30th USENIX Security Symposium (USENIX Security 21)},
year = {2021},
isbn = {978-1-939133-24-3},
pages = {91--108},
}

@inproceedings {308911,
author = {Lassak, Leona and Lindemann, Nicklas and Kowalewski, Marvin},
title = {From TOTPs to security keys: studying the reality of passwordless FIDO2 authentication with PIN and biometrics in a corporate environment},
year = {2025},
isbn = {978-1-939133-51-9},
publisher = {USENIX Association},
address = {USA},
booktitle = {Proceedings of the Twenty-First USENIX Conference on Usable Privacy and Security},
articleno = {21},
numpages = {19},
location = {Seattle, WA, USA},
series = {SOUPS '25}
}

@article{jannettstate,
author = {Louis Jannett and Andreas Mayer and Maximilian Westers and Vladislav Mladenov and Christian Mainka and J{\"o}rg Schwenk},
title = {The State of Passkeys: Studying the Adoption and Security of Passkeys on the Web},
booktitle = {the Proceedings of the 35th USENIX Security Symposium},
year = {2026},
address = {Baltimore, MD},
url = {https://www.usenix.org/conference/usenixsecurity26/presentation/jannett},
publisher = {USENIX Association},
month = aug
}

@inproceedings{dietz2012origin,
	title={$\{$Origin-Bound$\}$ Certificates: A Fresh Approach to Strong Client Authentication for the Web},
	author={Dietz, Michael and Czeskis, Alexei and Balfanz, Dirk and Wallach, Dan S.},
	booktitle={Proceedings of the 21st USENIX Security Symposium},
	pages={317--331},
	year={2012}
}

@inproceedings{gerlitz2023adventures,
	title={Adventures in recovery land: Testing the account recovery of popular websites when the second factor is lost},
	author={Gerlitz, Eva and H{\"a}ring, Maximilian and M{\"a}dler, Charlotte Theresa and Smith, Matthew and Tiefenau, Christian},
	booktitle={Nineteenth Symposium on Usable Privacy and Security (SOUPS 2023)},
	pages={227--243},
	year={2023}
}

@inproceedings{gilsenan2023security,
	title={Security and privacy failures in popular $\{$2FA$\}$ apps},
	author={Gilsenan, Conor and Shakir, Fuzail and Alomar, Noura and Egelman, Serge},
	booktitle={Proceedings of the 32nd USENIX Security Symposium},
	pages={2079--2096},
	year={2023}
}
